\documentclass{aa}
\usepackage[T1]{fontenc}
\usepackage[latin1]{inputenc}
\usepackage{txfonts}
\usepackage{graphicx}
\usepackage{url}

\usepackage{natbib}
\bibpunct{(}{)}{;}{a}{}{,} 

\def\sun{\hbox{$_\odot$}}

\begin{document}

\title{Solar off-limb line widths: Alfvén waves, ion-cyclotron waves, and preferential heating}
\titlerunning{Solar off-limb line widths}

\author{L. Dolla\inst{1} \and J. Solomon\inst{1,2}}

\offprints{L. Dolla}

\institute{Institut d'Astrophysique Spatiale, UMR8617, Univ Paris-Sud, Orsay, F-91405
   \and CNRS, Orsay, F-91405 \\
\email{laurent.dolla@ias.u-psud.fr; jacques.solomon@ias.u-psud.fr}
           }

\date{Received 16 May 2007 / Accepted 18 January 2008}

\abstract
   {Alfvén waves and ion-cyclotron absorption of high-frequency waves are frequently brought into models devoted to coronal heating and fast solar-wind acceleration. Signatures of ion-cyclotron resonance have already been observed \emph{in situ} in the solar wind (HELIOS spacecrafts) and, recently, in the upper corona (UVCS/SOHO remote-sensing results). } 
   {We propose a method to constrain both the Alfvén wave amplitude and the preferential heating induced by ion-cyclotron resonance, above a partially developed polar coronal hole observed with the SUMER/SOHO spectrometer. }
   {The instrumental stray light contribution is first substracted from the spectra. By supposing that the non-thermal velocity is related to the Alfvén wave amplitude, it is constrained through a density diagnostic and the gradient of the width of the \ion{Mg}{x} 625~\AA{} line. The temperatures of several coronal ions, as functions of the distance above the limb, are then determined by substracting the non-thermal component to the observed line widths. }
   {The effect of stray light explains the apparent decrease with height in the width of several spectral lines, this decrease usually starting about 0.1--0.2~R\sun{} above the limb. This result rules out any direct evidence of damping of the Alfvén waves, often suggested by other authors. We also find that the ions with the smallest charge-to-mass ratios are the hottest ones at a fixed altitude and that they are subject to a stronger heating, as compared to the others,  between 57\arcsec{} and 102\arcsec{} above the limb. This constitutes a serious clue to ion-cyclotron preferential heating. }
   {}


\maketitle
%
\section{Introduction}   
\label{sec Ion cyclotron resonance and Alfvén waves} 
%
Ion-cyclotron resonance is often suggested as one of the possible processes that accelerate the fast solar wind and heat the corona. 
Different modelling approaches are used, like hybrid simulations \citep[e.g.][]{Liewer99,Ofman04}, kinetic models \citep{Isenberg01,Isenberg01b, Marsch01, Vocks02b}, or fluid models. In this last category \citet{Cranmer99b} for example, include several ion species and show that it is necessary to regenerate the cyclotron waves through the wind to provide enough energy to the wind. This may be done, for example, by including an Alfvén wave turbulent cascade, as in \citet{Tu97} (two fluids) or \citet{Hu00} (four fluids), while \citet{Markovskii01} and \citet{Markovskii02} study the direct generation of cyclotron waves via plasma instabilities. 
 
In the models that use turbulent cascade, the Alfvén wave power is distributed into a power law spectrum, with ion-cyclotron waves being its high frequency limit. On the one hand, these waves are gradually absorbed by the particles as the different local ion gyrofrequencies decrease with altitude and match the wave frequency, taking into account the Doppler effect due to the particle velocity \citep[for the theoretical aspects of the cyclotron absorption, see e.g.][]{Hollweg02}.  On the other hand, they replenish through the energy cascade process from low to high frequencies. 
The large decrease in the magnetic field intensity because of the rapid expansion of the flux tubes can favour the resonance occurring at a relatively low altitude above the solar surface, as in the so-called "funnels" \citep[e.g.][]{Hackenberg00,Li02}. In this respect, the superradial expansion of the flux tubes in  coronal holes \citep{Kopp76}, which are thought to be the source of the fast solar wind, is also important. 

In some models, Alfvén waves are directly dissipated to provide energy to the wind. 
According to \citet{Parker91}, though, this dissipation is not significant below 5~R\sun{}, so that it is more likely that Alfvén waves stay undamped in the low corona. 

Several signatures of wave-particle interaction, which can be attributed to the ion-cyclotron resonance, have already been observed in the solar wind. The \emph{in situ} observations of the Helios spacecrafts, for example, showed differential speed of minor heavy ions, anisotropies of temperatures ($T_\perp / T_\parallel$), and preferential heating of some ion species \citep{Schwenn91}. In the upper corona, the UVCS spectro-coronometer on board SOHO made it possible to observe the same kinds of signatures from 2.5 to more than 5~R\sun{} from disc centre, through spectroscopic studies coupled to some modelling \citep{Li98, Kohl98, Cranmer99, Esser99}. 
In the lower corona, in contrast, there is lack of conclusive studies, 
and yet it is particularly important to verify whether any process that accelerates the fast solar wind starts acting in the lower corona. Observing this region is also important for constraining the energy actually dissipated by the cyclotron resonance and for differentiating  the different theories that have been proposed. 

In this article, we concentrate on the possibility of revealing the preferential heating of the ion species having the lowest charge-to-mass ratios ($q/m$). 
Retrieving the temperature of different ion species can be made via the analysis of the widths of emission lines. It is worthwhile noting that the temperature diagnostics based on line ratios are sensitive to the electron temperature and cannot help in measuring the ion one. 

Our paper is organised as follows. In Sect.~\ref{sec Retrieving the information}, we present the usual interpretation of the coronal line widths and of their variations above the limb, as well as solutions proposed in the literature for distinguishing the thermal and the non-thermal components. 
In Sect.~\ref{sec Observation and data processing}, we present the observations and the associated data processing, with emphasis on the correction for the stray light contribution in the spectra. In Sect.~\ref{sec method separation of contributions}, we explain the method we used to separate thermal and non-thermal Doppler effects in the measured line widths. The results are presented in Sect.~\ref{sec results}, while Sect.~\ref{sec conclusion} gives the conclusions. 
%
\section{Retrieving the information contained in coronal line widths}   
\label{sec Retrieving the information}
%
\subsection{Interpretation of the VUV line widths of coronal ions} 
The plasma observed above the limb is optically thin for most of the transition region and coronal lines in the VUV range observed by instruments like SUMER \citep{Dere93, Chae98, Lee00}. The two main contributions to the line width then come from thermal and non-thermal Doppler effects. 
The non-thermal velocity, also known as "unresolved" velocity, $\xi$ was originally introduced to interpret the  width of the transition region and coronal lines observed on the disc. In fact, there is an excess broadening as compared to the thermal component due to the formation temperature $T_\textrm{f}$ of these lines.  
This additional contribution was first attributed to velocity fluctuations induced by acoustic waves \citep{Boland73}, before MHD waves or fluid turbulence were considered \citep[e.g.][]{Boland75, Hassler90, Doyle99, Harrison02}. The term "unresolved velocity" in fact refers to motions on scales smaller than the resolution scale. One should also take into account the motions integrated on the depth of the line of sight (LOS), and on scales smaller than the temporal resolution of the instrument. 
If both thermal and non-thermal contributions are Gaussian \citep{Tu98}, the half-width $\sigma$ at $1/\sqrt{e}$ follows the formula given by \citet{Dere93}: 
\begin{equation}  \label{eq sigma}
   \sigma^2=\frac{\lambda^2}{2c^2} (\frac{2kT}{m}+\xi^2) 
\end{equation}
where $\lambda$ is the wavelength, $c$ the speed of light, $k$ the Boltzmann constant. 
The non-thermal velocity $\xi$ is proportional to the mean velocity fluctuations integrated on the observed solid angle and the exposure time,  $<\delta v^2>^{1/2}$. The coefficient of proportionality, relatively close to 1, depends on the degrees of freedom of the fluid motion \citep[e.g.][]{Tu98}. 

Variations in the non-thermal velocity have been investigated by several authors: (i) centre-to-limb variations by \citet{Mariska78}, \citet{Chae98}, \citet{Lee00}, \citet{Doyle00}, see also \citet{Erdelyi98} for the interpretation in terms of MHD modes; (ii) variation depending on the nature of the observed region (coronal holes, "quiet Sun", active region); and (iii) radial variation out of the limb, which is of interest for this article and which we will present in the next section. Depending on all these different conditions, the value of $\xi$ varies roughly from 10 to $30~\textrm{km} \, \textrm{s}^{-1}$. 

From on-disc observations, we retain the results of \citet{Chae98}, which assumed that the temperature of each ion species that they observed was the corresponding formation temperature ($T_\textrm{f}$). They consequently deduced the non-thermal velocity from the line widths, and plotted it as a function of $T_\textrm{f}$. The same kind of analysis was also performed by, e.g., \citet{Boland75} and \citet{Dere93}. The curve of \citet{Chae98} increases from $\approx 5~\textrm{km} \, \textrm{s}^{-1}$ at $10^4~\textrm{K}$ up to $\approx 30~\textrm{km} \, \textrm{s}^{-1}$ around $3 \times 10^5~\textrm{K}$ (\ion{O}{vi}), then decreases to $\approx 17~\textrm{km} \, \textrm{s}^{-1}$ at $10^6~\textrm{K}$ (\ion{Fe}{xii}). 
One can roughly interpret this curve in terms of LOS integration: different $T_\textrm{f}$ of ions observed at the same pointing on the disc can be associated to different altitudes above the solar surface. The curve then traces the value of the non-thermal velocity through the different temperature layers of the solar atmosphere \citep{Mariska78}. 
%
\subsection{Past observations of line widths above the solar limb} 
\label{sec past observations}
%
Most authors who have observed above the limb, in coronal holes, report that the line widths first increase with the altitude before presenting a plateau or, sometimes, starting to decrease at 1.1~R\sun{} from disc centre: e.g. \citet{Hassler90}, with the \ion{Mg}{x} 609 and 625~\AA{} lines up to 1.2~R\sun, or \citet{Banerjee98}, with SUMER/SOHO observing the \ion{Si}{viii} 1445~\AA{} line. 
These authors interpreted this behaviour as caused by Alfvén waves, with the increase due to flux conservation and decrease due to damping. 
Using CDS/SOHO, \citet{O'shea03} observed the width of the \ion{Mg}{x} 625~\AA{} line starting to decrease around 1.06~R\sun, the exact position varying with the data sets, but always for the same value of the line widths. 
But \citet{O'Shea04}, when observing the \ion{Si}{xii} 520~\AA{} line in addition to the two \ion{Mg}{x} 609 and 625~\AA{}  lines, found a discrepancy between both species and concluded that the behaviour of the \ion{Mg}{x} lines results from photoexcitation growing as the altitude increases. 
Still above a coronal hole, using SUMER, \citet{Lee00} found a link between the altitude where the line widths change their behaviour and the intensity scale height. Nevertheless, they used a lot of particularly cold lines, which they could not observe more than 20\arcsec{} above the limb. They also found a curve $\xi = f(T_\textrm{f})$ as did \citet{Chae98}. 

Above a quiet Sun region at the equator, \citet{Harrison02} observed a decrease in the line width of the \ion{Mg}{x} 625~\AA{} line, with CDS, starting at about 1.04~R\sun{}, the width being narrower than above coronal holes. They linked this behaviour to the damping of Alfvén waves in close field loops. \citet{Wilhelm04}, using SUMER still at the equator but at a different date, did not find the same behaviour. The results of both instruments were reconciled after joint observations \citep{Wilhelm05} that revealed no decrease, but a slight plateau. Besides, \citet{Peter03} observed a local maximum of the widths of transition region lines (\ion{O}{v} and \ion{S}{vi}), less than 10\arcsec{} ($\approx 0.01~\textrm{R\sun}$) above the limb, which they attribute to the dissipation of ion-cyclotron waves, the decrease at higher altitudes being due to collisional cooling when the ion-cyclotron dissipation vanishes. 

\citet{Singh03,Singh03b} analysed ground-based observations of \ion{Fe}{x}, \ion{Fe}{xiii}, and \ion{Fe}{xiv} lines (visible and infrared), up to 140\arcsec{} in streamers. On average, the \ion{Fe}{x} 6374~\AA{} line width increases with altitude, those of the \ion{Fe}{xiii} 10747 and 10798~\AA{} lines increase sligthly less, whereas that of the  \ion{Fe}{xiv} 5303~\AA{} line decreases slightly. Concluding that the line-width behaviour depends on the formation temperature, 
they reject the interpretation in terms of variation of Alfvén wave amplitude and prefer to explain it by an LOS effect in a thermally stratified corona. 
Still from the ground, \citet{Contesse04} observe no clear decrease in the width of the \ion{Fe}{xiv} 5303~\AA{} line, but instead a plateau up to 1.5~R\sun{} at the equator (east and west) and up to 1.2~R\sun{} above both poles, the observed line widths being larger above the poles than  above the equatorial regions.

Using UVCS/SOHO, it is possible to observe several solar radii above the solar limb, essentially the \ion{H}{i} Ly~$\alpha$ and the \ion{O}{vi} lines at 1032 and 1037~\AA{}. The observations showed that widths generally increase more with the altitude in the upper part of the corona, i.e. above 2~R\sun{} from disc centre, than in the lower one \citep{Li98, Kohl98, Cranmer99, Esser99, Frazin99, Doyle99, Morgan04b}. 
%
\subsection{Distinguishing thermal and non-thermal Doppler effects} 
\label{sec distinguishing thermal/nonthermal effect}
The line width contains information about the result of potential heatings (the temperature $T_\textrm{i}$ of the different ions species) and about the possible source of heating, that is to say, the Alfvén wave amplitude via direct dissipation or via ion-cyclotron resonance after a turbulent cascade. All difficulties come from these two pieces of information merged in only one observable. Most methods used in the literature consist in making an assumption about one of the above quantities, depending on the purpose of the authors, i.e. constraining the Alfvén wave amplitude or revealing any preferential heating. 

One can find mostly two kinds of hypotheses on the ion temperature: the isothermal or the formation temperature hypotheses. One example of isothermal hypothesis can be found in \citet{Feldman00} or \citet{Doschek01}. They determine the electron temperature through the emission measure and assume that all species are in thermal equilibrium. Then, they find different values of $\xi$ for the different ion species and discuss about possible errors to explain why they do not obtain a unique value for $\xi$. In fact, \citet{Moran03} shows for a coronal hole that it is not possible for all ion species to have the same temperature and the same non-thermal velocity at the same time. 

Numerous authors prefer to assume that the temperature $T_\textrm{i}$ of each ion is the formation temperature of the line \citep[e.g.][]{Hassler90,Contesse04}. But this supposes ionization equilibrium, while several authors think that the plasma above coronal holes is weakly collisional \citep{Tu98,Banerjee98}. When a preferential heating process, like ion-cyclotron resonance, is at work on time scales smaller than the collisional cooling (cf. Table~\ref{tab characteristic times}), the assumption $T_\textrm{i} = T_\textrm{f}$ no longer makes sense.  

\citet{Doyle99}, for a coronal hole, use a more elaborated hypothesis for the temperature. They suppose that the electron temperature $T_\textrm{e}$ is equal to the formation temperature of the observed \ion{Si}{viii} line and deduce the proton temperature $T_\textrm{p}$ from hydrostatic equilibrium, constrained by a density diagnostic. Then, they suppose that $T_\textrm{\ion{Si}{viii}} = T_\textrm{p}$ (and $T_\textrm{p} > T_\textrm{e}$) before they deduce $\xi$ from the \ion{Si}{viii} line width. 

The first hypothesis about the non-thermal velocity $\xi$ that can be made is that all ion species have the same $\xi$. This is reasonable if this velocity comes from fluid motions, i.e. 
regardless of the species; this may be false if some species have high differential speeds \citep{Hollweg02}, which should not be the case in the lower corona. The problem is then to determine the value of $\xi$. 
\citet{Seely97} adjust $\xi$ so that the temperature of the Fe ion species, i.e. the heaviest element they observe, is similar to that of ion species of neighbouring $T_\textrm{f}$, as they suppose that all ion species that have a common $T_\textrm{f}$ are in thermal equilibrium. 
They then deduce the temperature of all the other ion species, which appear to be larger than their own formation temperature; the lower the formation temperature, the larger the difference. These authors  conclude that this situation may come from out-of-equilibrium  processes (heating or cooling phase) or from preferential heating. They also conclude that the underestimation of the ion temperatures led \citet{Hassler90} to overestimate $\xi$ at altitudes of $100-200~\textrm{arcsec}$. 

When looking for the signature of preferential heating in coronal holes, \citet{Tu98} cautiously give an upper and a lower limit to $\xi$. The lower one being 0, they set the upper one by setting to zero the thermal width of the lines from the heaviest ion having the smallest Doppler width (\ion{Fe}{xii}) in their data set. Then, they constrain the temperature of all ion species, and note a stronger heating of ions having the smallest $q/m$ between about 1.04 and 1.18~R\sun{}. 

One possibility for better constraining $\xi$ is to suppose that the non-thermal velocity is solely due to undamped Alfvén waves. To conserve the wave energy in an expanding flux tube, while the mass density $\rho$ decreases, the wave amplitude must increase so that \citep[e.g.][]{Moran01}  
\begin{equation}  \label{eq xi fn density}
 \xi \propto \rho^{-1/4}. 
\end{equation}
The line-width variation with the altitude has been analysed with respect to this effect in \citet{Hassler90} for \ion{Mg}{x} and \citet{Banerjee98} and \citet{Pekunlu02} for \ion{Si}{viii}. 

Conclusions that can be drawn from the study of VUV solar line widths appear to strongly depend on the method used to separate the thermal from the non-thermal contribution. Two points emerge from past studies: (i) the probable existence of a non-thermal velocity whose radial behaviour above the limb mimics that of undamped Alfvén wave amplitude. The question of whether these waves begin to damp at a certain point above the solar surface does not seem to have been solved. (ii) Another process may be at work that affects different lines in different ways. A $q/m$ dependence is suspected that could be associated to ion-cyclotron resonance. 

The purpose of this article is to verify and constrain both of these two points without using hypotheses that \emph{a priori} preclude one of them. We will focus on observations of a polar coronal hole, below 1.2~R\sun{}. To avoid bias resulting from the line formation (e.g. photoexcitation) and from unknown line blendings, it seems useful to observe as many lines as possible. A wide range of masses and $q/m$ of the ion species also favour the distinction between the relative contribution of thermal and non-thermal velocity in the observed line widths. 
\begin{table}[tbc]
	\caption{Comparison of the characteristic cyclotron time of some species with their heat exchange time with protons, in the lower corona. } 
	\centering
		\begin{tabular}{l c c c }
			\hline
			\hline
			Species & $q/m^{\mathrm{a}}$ & $\tau_c$ (s)$^{\mathrm{b}}$ & 
			$\tau^\textrm{\tiny E}$ (s)$^{\mathrm{c}}$ \\
			\hline
		  e$^{-}$ & 1758 & $4 \times 10^{-7}$ & 120 \\
			p$^{+}$ & 1 & $6 \times 10^{-4}$ & 3 \\
			\ion{Mg}{x} & 0.37 & $2 \times 10^{-3}$ & 1 \\
      \ion{Fe}{x} & 0.16 & $4 \times 10^{-3}$ & 2 \\
			\hline
		\end{tabular}
  \begin{list}{}{}
  \item[$^{\mathrm{a}}$] Charge-to-mass ratio, normalized to the proton one. \ion{Mg}{x} and \ion{Fe}{x} are the ion species having the highest and lowest values of $q/m$, respectively, in the spectrum observed by SUMER.
  \item[$^{\mathrm{b}}$] $\tau_c = 2 \pi \frac{m}{q \, B}$, with $q$ the charge and $m$ the mass of the particle, and $B$ the magnetic field. 
  \item[$^{\mathrm{c}}$] Heat exchange time with protons $\tau_\textrm{\scriptsize ep}^\textrm{\tiny E}$ and  $\tau_\textrm{\scriptsize ip}^\textrm{\tiny E}$ \citep[e.g.][p.~68]{Wesson04}. 
  
All these times are calculated with the following conditions: $B=1~\textrm{gauss}$, electron density  $n_\textrm{e}=10^{14}~\textrm{m}^{-3}$, and proton and electron temperatures both equal to $10^6$~K ($\beta \approx 0.35$). 
  \end{list}
	\label{tab characteristic times}
\end{table}
%
\section{Observations and data processing} \label{sec Observation and data processing}
%
The VUV Spectrometer SUMER on board the SOHO spacecraft is described in \citet{Wilhelm95}, \citet{Wilhelm97}, and \citet{Lemaire97}. 
Table~\ref{tab data sets} briefly presents all data sets used in this article. The main one, called set~1, consists of ten spectral ranges, observed using detector A on 30 May 2002 during MEDOC Campaign~\#9. 
The lines of interest are listed in Table~\ref{tab results data set 1}. The complete observation was about 13 hours long. The centre of the slit was pointed at X=0\arcsec{}, Y=1150\arcsec{} (i.e. at the North Pole). The $1\arcsec \times 300\arcsec$ slit was used, oriented in the north-south direction, covering a wide range of altitudes at the same time (from about 50\arcsec{} to 350\arcsec{} above the solar limb, with an apparent solar radius of 956\arcsec). The width of 1\arcsec{} reaches a compromise between the photon statistics and the instrumental width. 
\begin{figure}
	\centering
		\includegraphics[width=0.45\textwidth]{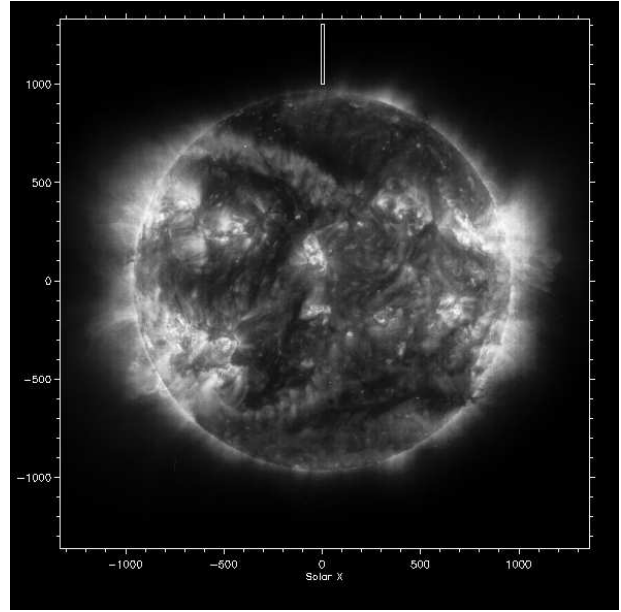}
	\caption{EIT/SOHO context image for data set~1 (195~\AA{} channel). The spatial coverage of the SUMER slit is shown. }
	\label{fig pointing_May2002}
\end{figure}
\begin{table}
	\caption{Description of data sets. } 
	\centering
		\begin{tabular}{c c c c}
			\hline
			\hline
      Data set$^{\mathrm{a}}$ &  Date   &  Position$^{\mathrm{b}}$  &  R\sun{} (arcsec)$^{\mathrm{c}}$ \\
			\hline
			 1         &   30 May 2002   &  X=0, Y=1150  &  956.5   \\      
			 2         &   28 May 2002   &  X=0, Y=-1150 &  956.5   \\  
       3         &   19 Nov 2003   & X=0, Y=1145   &  980.0   \\  
       4         &   12 June 2004  & X=0, Y=-1140  &  960.0   \\   
			\hline  
		\end{tabular}
  \begin{list}{}{}
  \item[$^{\mathrm{a}}$] Observations are all made with the $1\arcsec \times 300\arcsec$ slit, and detector A. In data sets 2, 3, and 4, only the \ion{Mg}{x} 625~\AA{} line was analysed in the present paper, for comparison purpose. 
  \item[$^{\mathrm{b}}$] Position of the centre of the slit, in arcseconds from disc centre. 
  \item[$^{\mathrm{c}}$] Solar radii as seen from SOHO. 
  \end{list}
	\label{tab data sets}
\end{table}

\begin{figure}
	\centering
		\includegraphics[width=\linewidth]{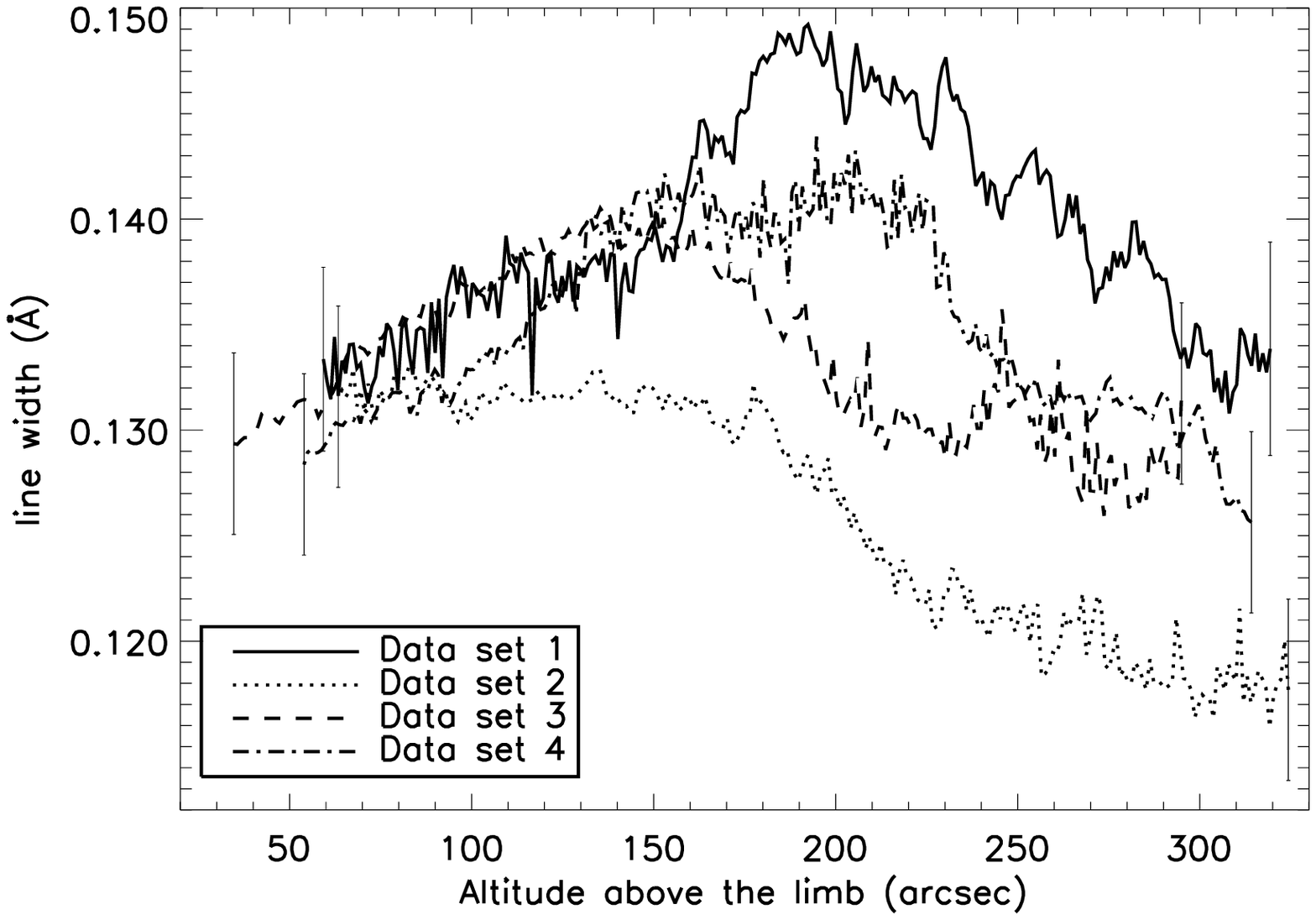}
		\includegraphics[width=\linewidth]{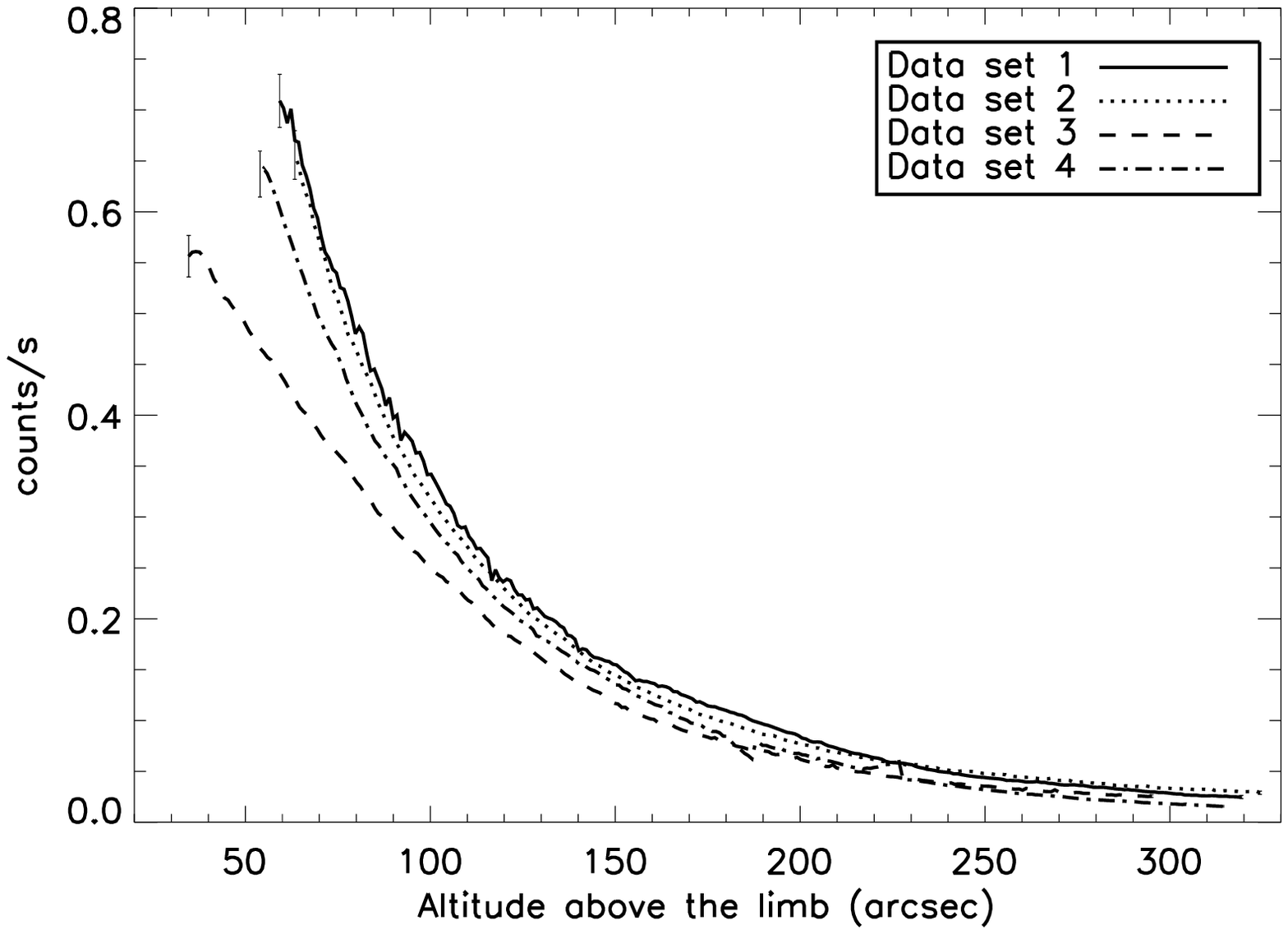}
	\caption{Comparison of the line width (top panel) and the total counts per second (bottom panel) measured for the \ion{Mg}{x} 625~\AA{} line (observed in second order; the measured width is then twice as large as the width that would be measured in first order), for the four data sets acquired above the solar poles (Table~\ref{tab data sets}). The error bars are only plotted for both extremities in the abscissas, for better visibility. The stray light contribution has not been removed yet. The total counts in the line are discussed in Sect.~\ref{sec No evidence for damping of Alfven waves}. }
	\label{fig Mg X different data sets}
\end{figure}

The EIT/SOHO images on that day (Fig.~\ref{fig pointing_May2002}) show that the coronal hole appearing on the solar disc at the North Pole is not developed well. Inspection of EIT movies of the preceding and following solar rotations reveals that the coronal hole is far from being symmetric, and highly elongated along one particular direction towards the equator. Nonetheless, it appears that some part of the coronal hole is lying right under the position we are pointing at in the plane of the sky, and LASCO images show no streamer on the north-south axis. However, we cannot dismiss the possibility of a contamination of the LOS by any plasma belonging to non-hole (i.e. "Quiet Sun", QS) part of the corona. 
To gain more confidence, we compared the behaviour of the width of the \ion{Mg}{x} 625~\AA{} line (observed at 1249.88~\AA{} in second order) in three other data sets.  
Data set~2 corresponds to the south pole observed two days before data set~1 (same altitude range); the EIT survey shows that there is no coronal hole at the south pole at this period. Data sets 3 and 4 correspond to other polar coronal holes (CH). Figure~\ref{fig Mg X different data sets} shows that the behaviour of the line width in set 1 compares well to those in sets 3 
and 4, because it keeps increasing at low altitudes and then decreases at the highest altitudes.  
In contrast, data set~2 displays a different behaviour, with a large plateau starting from  60\arcsec{}, and it will be analysed more thoroughly in a future paper. 
According to \citet{Wilhelm04}, the \ion{Mg}{x} 609 and 625~\AA{} line widths are larger in coronal holes than in the Quiet corona (by more than 30\% at 100\arcsec{}). Data sets 1 to 4 do not show such large differences. The maximum one is between sets 1 and 2, and the \ion{Mg}{x} line width in set 1 is only larger than that of set~2 by $\approx 4\%$ at 100\arcsec{}. In fact, from 60\arcsec{} to 100\arcsec{}, our values are higher than those of \citet{Wilhelm04} in the Quiet corona by about 15\% and lower than those in coronal holes by about 15\%. Nevertheless, they are closer to what was observed by \citet{Doschek01} in coronal holes, although a contamination by non-hole material cannot be dismissed, especially because the magnetic topology is far from being as simple as during solar minima. The density, discussed in Sect.~\ref{sec Density diagnostic}, is similar to what has been previously measured in coronal holes. 

A density diagnostic has been made using additional data acquired on 30 May 2002. 
The lines of \ion{Si}{viii} at 1440 and 1445~\AA{} were observed by SUMER, covering the same field of view as data set~1, with a $4\arcsec \times 300\arcsec$ slit. This wider slit thus provides better photon statistics for the line ratio, at the expense of an optimal instrumental width. The total exposure time was about 2 hours and 50 minutes. 

Raw data 
were corrected using the standard SUMER procedures provided by Solar Software (flatfield and geometric distorsion of the detector). Note that the observation of faint lines with long exposure times is only possible because the detector noise is low, $\approx 3 \times 10^{-5}~\textrm{counts\,s}^{-1}\,\textrm{px}^{-1}$ \citep{Wilhelm97}, which makes $\approx 0.1~\textrm{counts\,px}^{-1}$ for a 1-hour exposure time. The final spectra were obtained through the addition of consecutive 300 second exposures, after visual inspection of each exposure. The complete detector image was recovered ($1024 \times 360$~pixels, wavelength $\times$ spatial dimension Y). The instrument is stigmatic in the spatial dimension Y, with a theoretical spatial resolution around 1\arcsec{}. For better statistics, several spatial pixels were added before analysing the spectral profiles. These intervals appeared to be narrow enough to avoid any significant artificial broadening of the lines owing to the residual distortion of the detector. A rough estimate of the altitude corresponding to each spectrum was provided through the pointing coordinates of the instrument. 
%
\subsection{Correction from the instrumental stray light} 
\label{sec correction from the instrumental stray light}
\begin{figure}
	\centering
		\includegraphics[height=0.25\textheight]{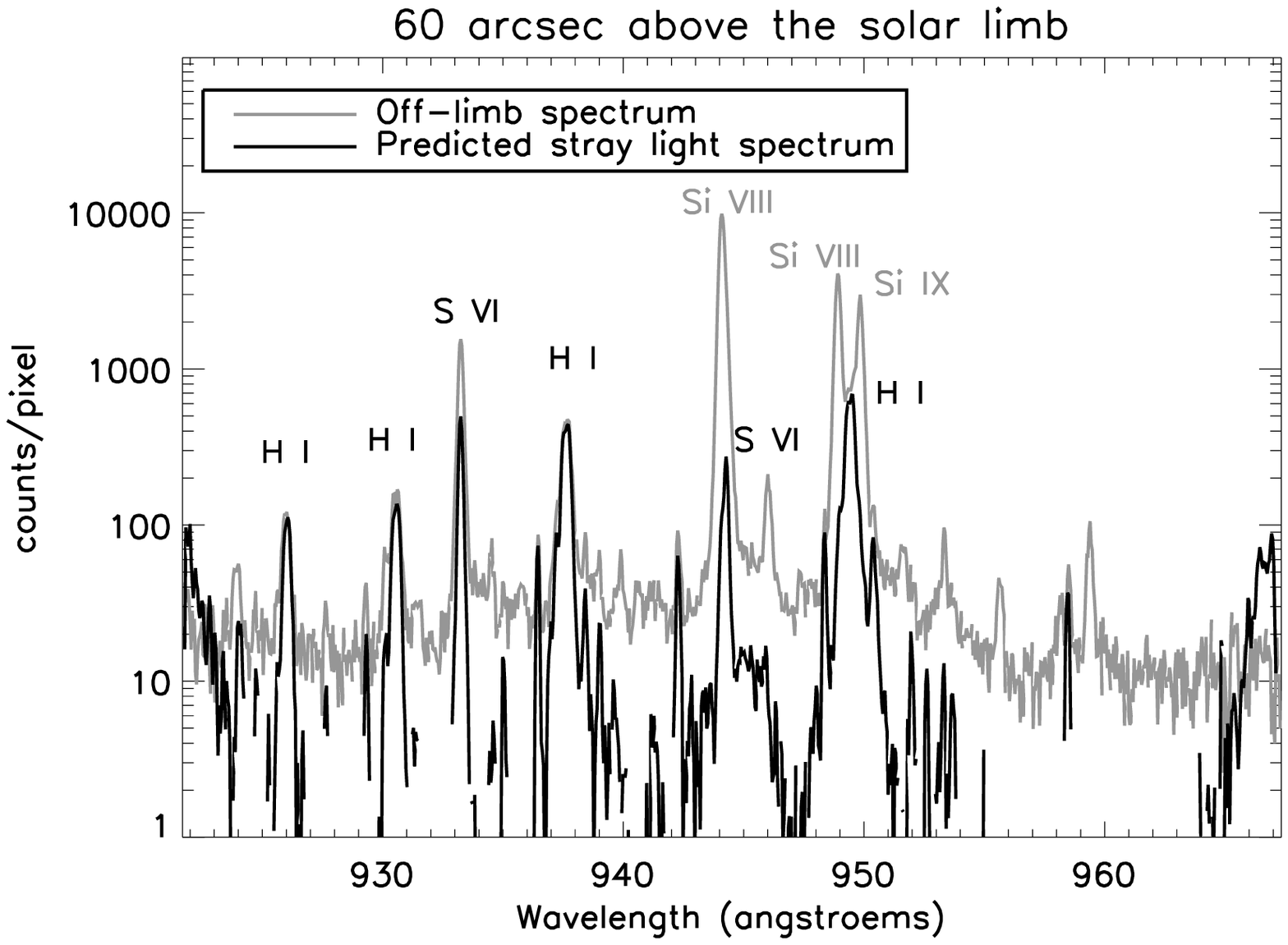}
		\includegraphics[height=0.25\textheight]{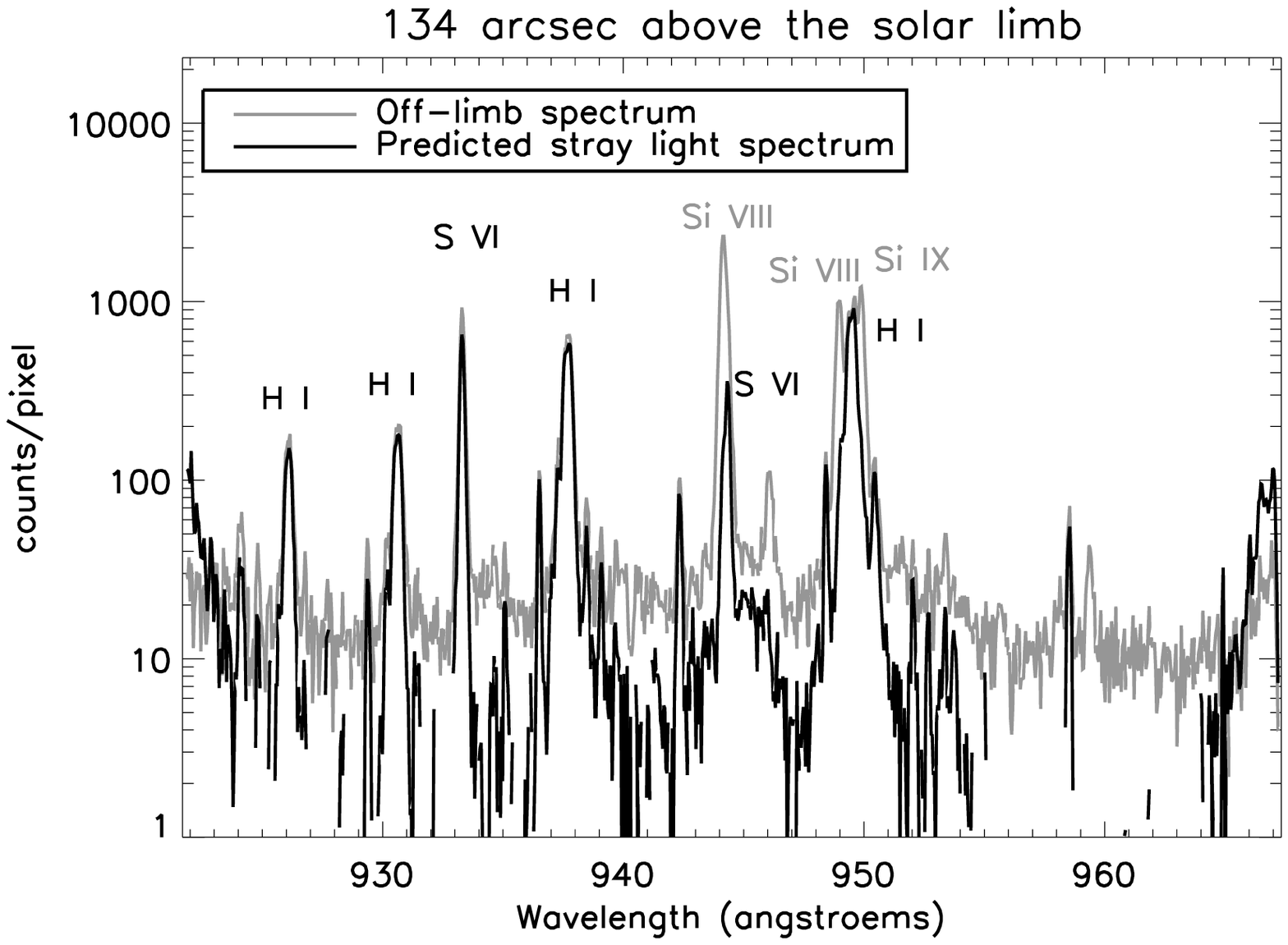}
		\includegraphics[height=0.25\textheight]{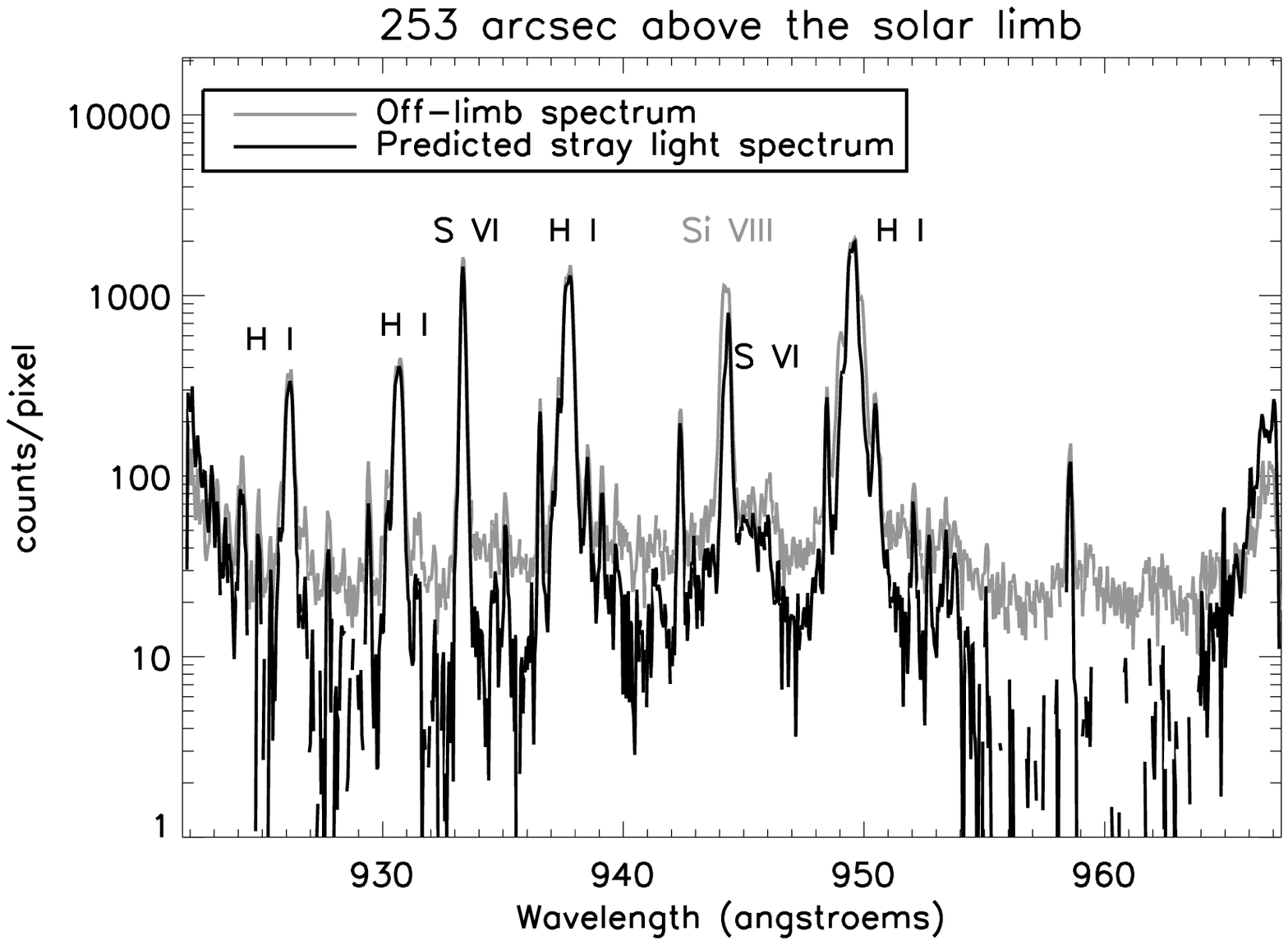}
	\caption{Examples of stray-light spectrum predictions for a spectral range covering $\approx 40$~\AA{} around the \ion{Si}{viii} line at 944~\AA{} (data set~2). Each observed off-limb spectrum (in grey) is the sum of an instrumental stray-light spectrum (predicted in black) and of the real coronal spectrum. As the altitude increases (here, from 60  to 134 and 253 arcsec above the limb), several cold lines become more prominent in the off-limb spectrum (like  \ion{H}{i} and \ion{S}{vi} lines, here). Meanwhile, some coronal lines become more and more contaminated by the blending with neighbouring cold lines (like \ion{Si}{viii} blended with \ion{S}{vi}); blending can also be caused by the same coronal line when this one is prominent in on-disc spectra (e.g., for \ion{Mg}{x} around 1250~\AA{} in second order). In these figures, spectra are in counts per wavelength pixels (with integration over wider spatial ranges for increasing altitude). } 
	\label{fig blending_SiVIII_944_ch16}
\end{figure}
As SUMER is not a coronagraph (no occultor is included), the solar disc produces stray light during off-limb observations, essentially via the point-spread function properties induced by defects in the reflective surfaces, of spatial sizes of the same order of magnitude as the observed wavelength. Note that, whatever the pointing is, the complete image of the Sun is permanently formed by the main parabola. 

This instrumental stray light is negligible for on-disc observations, but the contamination increases as the altitude of observation above the limb increases, until it is responsible for the quasi-complete photon rate detected above 1.6~R\sun{} \citep{Feldman99}.  This problem is even more critical in coronal holes than above the "quiet" corona, as the emissivity is lower. 
Consequently, each off-limb spectrum is the sum of the real spectrum emitted by the corona and of the stray-light spectrum. At each point of the image of the Sun produced by the mirror, the intensity of the stray light results from contributions convolved over the entire source (disc plus corona seen above the limb). But the main contribution comes from the disc, where the different  regions (active regions, coronal holes, and "Quiet Sun") emit different spectra. This results in a  stray-light spectrum that varies with the position and date of observations. 

Every line that can be observed on the disc contributes to the stray-light spectrum observed above the limb, so that coronal lines (\ion{Mg}{x} 625~\AA{}, \ion{Fe}{xii} 1242~\AA{}, etc.) are not only blended with their own stray light, but they can also be blended with colder lines of neighbouring wavelengths. This last effect is usually underestimated. 

To correct for the instrumental stray light, we used a method already described in \citet{Dolla03}, \citet{Dolla04}, and \citet{Dolla_PhD}, which enabled us to predict the stray-light spectrum contaminating the coronal spectrum observed at any altitude. This predicted spectrum is then substracted from the off-limb spectrum \citep[see also][]{Banerjee00, Singh03, Singh03b}. 
For the observed spectral ranges, a reference spectrum was taken well above the limb, so that the spectrum can be considered as purely due to stray light. All reference spectra were acquired with a long exposure (several hours), usually pointing at (X=0\arcsec{}, Y=1700\arcsec{}). The same slit was used as for the off-limb spectra. 
The stray-light spectrum is then predicted at any lower altitude by determining the factor that makes the total counts in a reference cold line coincide in both the observed off-limb spectrum and the predicted stray-light spectrum. If no line offers sufficient statistics, one can use the continuum, such as for the \ion{Fe}{x}/\ion{Fe}{xi} spectral range, near 1465~\AA{}. 
Some examples of stray light prediction are shown in Fig.~\ref{fig blending_SiVIII_944_ch16} \citep[see also][]{Dolla03,Dolla04,Dolla_PhD}. 

We showed in \citet{Dolla03} that reference spectra taken directly on the disc pointing at a "quiet Sun" region, and averaged over all the spatial pixels observed by the slit, do not allow a precise prediction of the stray-light spectrum, because these spectra do not take every kind of region on the disc into account. 
With high-altitude spectra, on the contrary, the convolution on the entire disc has already been performed. This also avoids modeling it like in \citet{Gabriel03}. Of course, the extrapolation of the high-altitude spectrum as a stray-light reference at any altitude is not totally correct, as the different contributions from the different areas of the disc are modified. This affects the ratios between the predicted lines, but the relative error on this ratio becomes more negligible as the altitude increases \citep[cf.][]{Feldman99, Doschek00}, precisely as the absolute correction increases. 

\begin{figure*}
	\centering
		\includegraphics[width=0.45\linewidth]{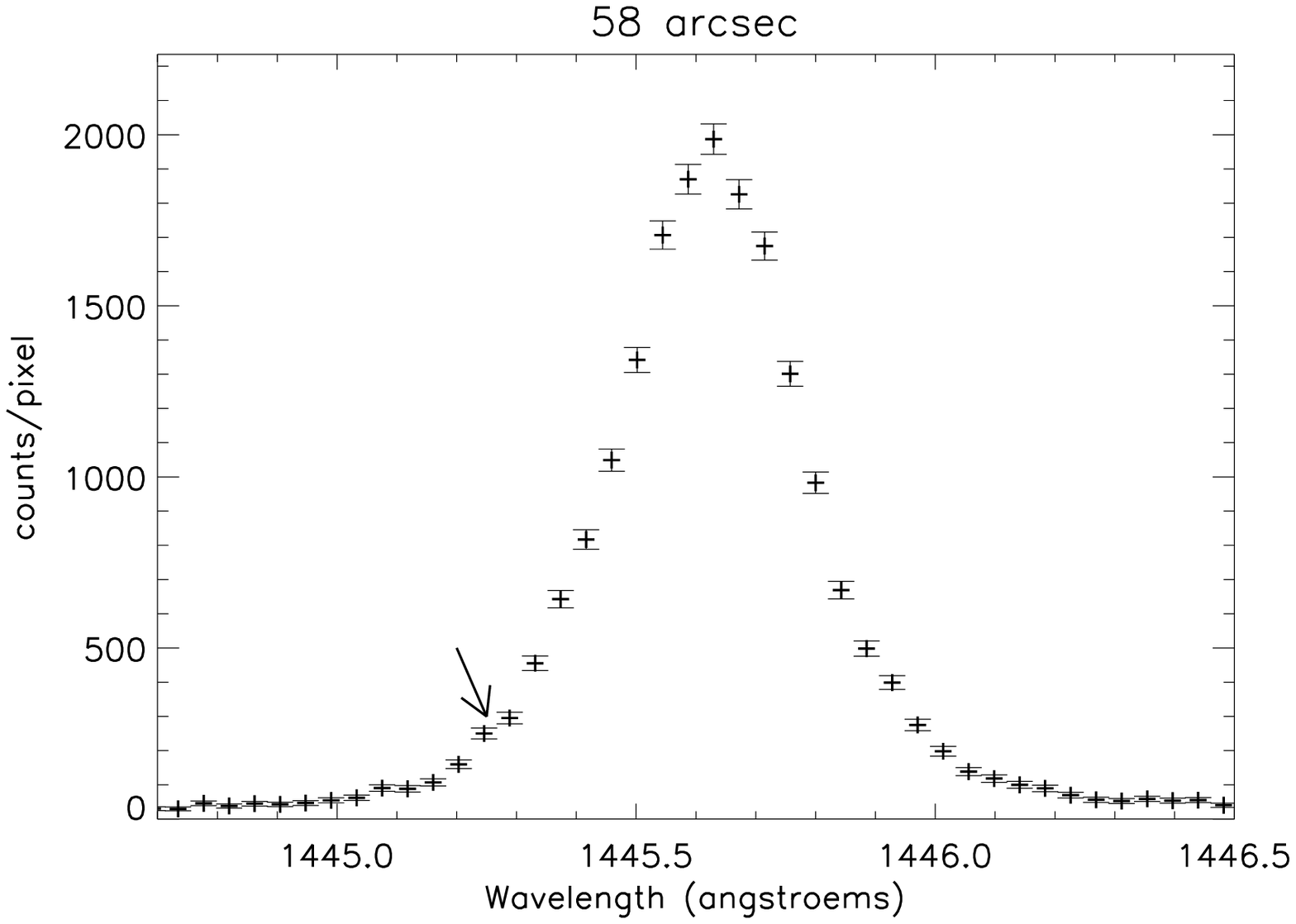}
		\includegraphics[width=0.45\linewidth]{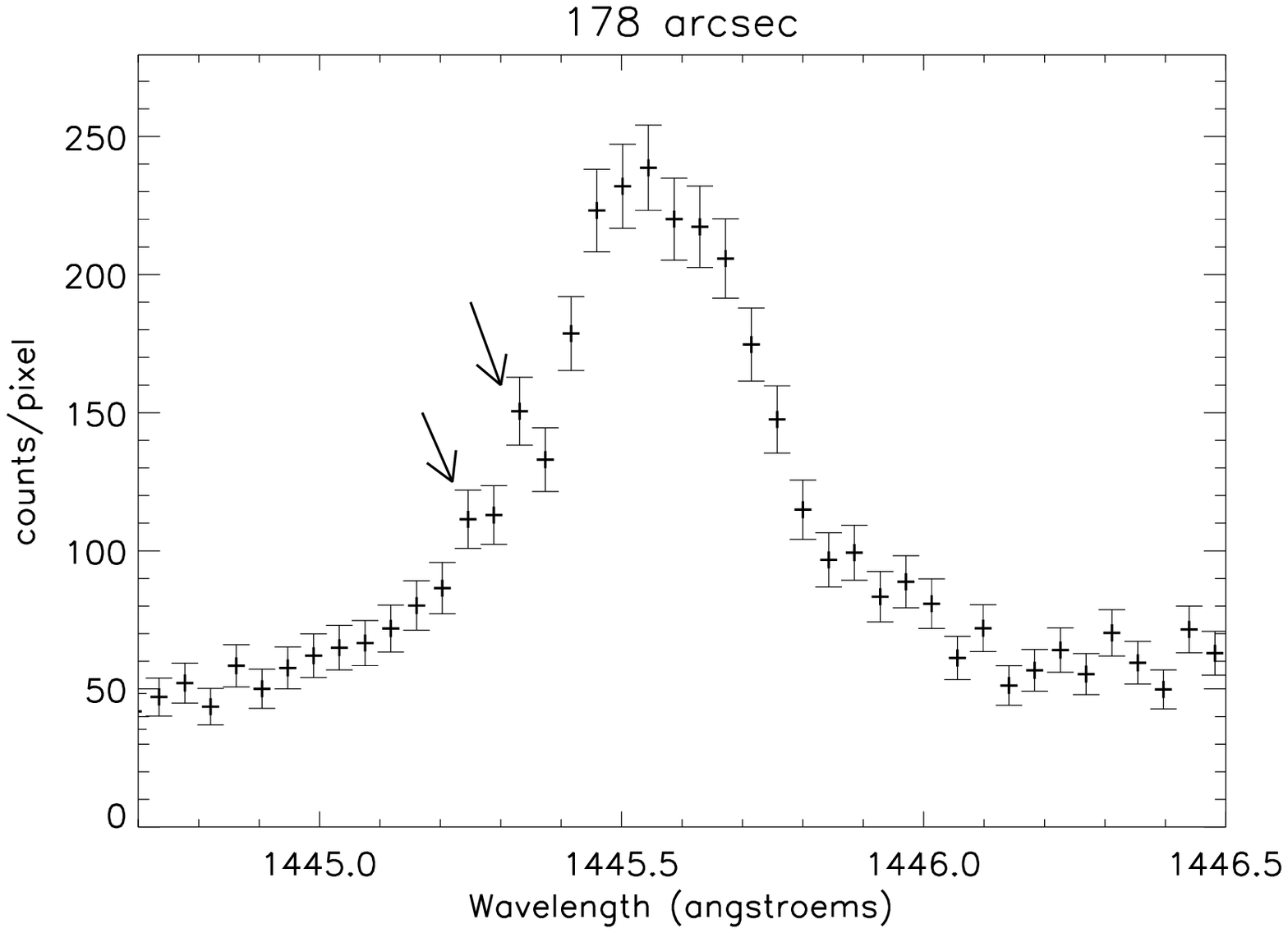}
	\caption{\ion{Si}{viii} line profiles (1445~\AA{}) at 58\arcsec{} and 178\arcsec{} above a coronal hole (data set~1). No correction of the stray light has been applied, because no high-altitude reference spectrum was taken for that. The "bumps" marked by the arrows, getting larger with the altitude and present on other data sets, suggest that this line is slightly contaminated by one cold line or more, appearing as instrumental stray light. }
	\label{fig SiVIII profiles}
\end{figure*}

During the observations, a high-altitude reference spectrum for the stray light was not taken for every spectral range observed at lower altitudes in the corona. In particular, judging from previous SUMER data and from spectral lines atlases, we thought that this was not necessary for the \ion{Si}{viii}  1440 and 1445~\AA{} lines. Nevertheless, Fig.~\ref{fig SiVIII profiles} shows a bump in the blue wing, growing as the altitude
increases. The same bump can be found in different data sets, so that statistical error is excluded. A possible stray light contamination, also noticed by \citet{Banerjee98}, may explain it. We suspect that another bump is present on the red wing. In fact, these bumps do not appear when using the 4\arcsec{} slit, probably because the wider instrumental profile smoothes the emission profiles. 

According to \citet{Ekberg03}, the \ion{Si}{xi} 604.15~\AA{} line is blended with an \ion{Fe}{vii} line. The blending line is a cold one \citep[class "b", following the designation of][]{Feldman97}, so that any direct blending is probably negligible above the limb. A stray light contamination is likely, but has been underestimated at the time of the data acquisition, and no high-altitude spectrum is available for the stray light 
determination.  
%
%
\subsection{Gaussian width} 
The half-width at $1/\sqrt{e}$ is provided through a least-square Gaussian fit including a second-order polynomial. The spectrum corrected for the stray light is in counts\,px$^{-1}$, so as to use statistical errors as weights. 
The width in pixels is turned into angstroms by using the known dispersion of the
spectrometer. The equivalent size of a pixel ranges from 42 to 45~m\AA{} in first order, depending on
the observed wavelength. 
No precise calibration of the line wavelength is necessary, and we used the values given by \citet{Feldman97}. The fitting procedure provides a 1-sigma error on the width, which we set to 0.1 pixel as a minimum. When the counts at line maximum were below 150 or so, the fitting procedure became unstable, so that we only show results of lines presenting more than 150 counts at their maximum.  

The spectrum observed on the detector corresponds to the convolution of the spectrum arriving at the aperture of the spectrometer with the instrumental profile. This profile mainly comes from the finite width of the slit. 
Both off-limb and stray-light reference spectra are acquired with the same slit width,   therefore the corrected spectrum is still convolved with the same instrumental profile  \citep{Dolla_PhD}. 

The instrumental width, for detector A, does not depend on the wavelength. The procedure con\_width\_funct\_3 in Solar Software\footnote{A newer procedure con\_width\_funct\_4 differs only for detector B. }, updated following a study made by \citet{Chae98}, remove it through a deconvolution process, which proves to be equivalent to removing (quadratically) a Gaussian contribution of 42~m\AA{} (99~m\AA{} in FWHM). 
%
%
\section{Method of separating the two contributions in the coronal line widths}
\label{sec method separation of contributions}
As shown in \citet{Dolla04}, that the non-thermal velocity is not well-constrained prevents  direct conclusion about the preferential heating, by plotting either the temperature at a given altitude or the difference in temperature between two altitudes, as a function of $q/m$. In effect, owing to the small number of available lines, the ion species that are more likely to take advantage of cyclotron heating, i.e. having low $q/m$ ratios, are also those with low $1/m$ ratios, that is to say, those whose width is more sensitive to the value of $\xi$ (cf. Eq.~\ref{eq sigma}). 

Instead of making an assumption on the value of $\xi$, we preferred to make an assumption on its nature. If we suppose that Alfvén waves are responsible for the non-thermal velocity, then its radial variation can be constrained using Eq.~\ref{eq xi fn density} and a density diagnostic. We also consider that the \ion{Mg}{x} species, having the highest $q/m$ ratio in our data set, is less likely to be heated through the cyclotron resonance, and we suppose that its line width will only increase because of an increase in the non-thermal velocity. This hypothesis will then be verified \emph{a posteriori} on a given interval. 
%
%
\subsection{Density diagnostic} \label{sec Density diagnostic}
\begin{figure}
	\centering
		\includegraphics[width=\linewidth]{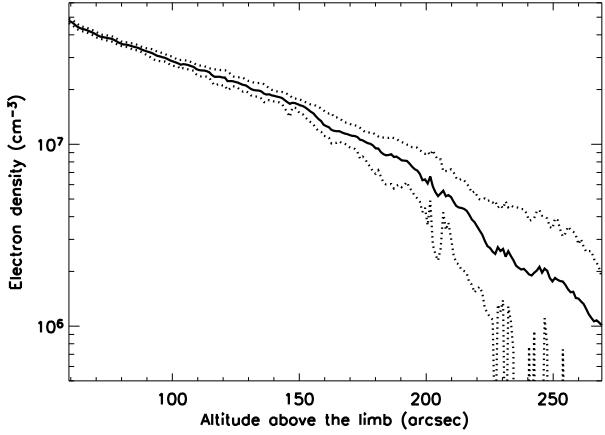}
	\caption{Electron density n$_\textrm{e}$ with increasing altitude above the North polar coronal hole (data set~1). The dotted lines denote the error bars. }
	\label{fig density}
\end{figure}

The \ion{Si}{viii} spectra are produced by a running sum over 40 pixels in the spatial direction of  the detector. 
The total counts in the spectral ranges corresponding to both \ion{Si}{viii} lines (1440 and 1445~\AA) are integrated. 
The continuum contribution is then substracted. 
The mass density $\rho$ can be obtained through the electron density n$_\textrm{e}$, which is derived in Fig.~\ref{fig density} from the theoretical curve linking the line ratio (1445/1440) to the electron density \citep[][their Fig.~1]{Doschek97}. 

Above 270\arcsec{}, the measured line ratio becomes lower than 1, so that no solution is given by the diagnostic. The \ion{Si}{viii} line ratio becomes far less sensitive at electron densities below $10^{7}~\textrm{cm}^{-3}$ and the 1445~\AA{} line suffer from the stray light mentioned above. For all these reasons, we have less confidence in the results above 180\arcsec{}. 

These density values are similar to those of \citet{Doschek97} measured in a polar coronal hole and at the same altitude range. They nevertheless decrease more rapidly with the altitude. 
%
%
\subsection{Determination of $\xi$ at the lowest altitude} 
Equation~\ref{eq sigma} can be rewritten as the quadratic sum of thermal and non-thermal velocity :
\begin{equation} \label{eq v2}
   v^2 = \frac{2kT}{m}+\xi^2. 
\end{equation}
By using the assumption that the \ion{Mg}{x} 625~\AA{} line-width variation is solely due to a variation in $\xi$ between positions $r_0$ and $r$ (i.e. $\Delta T(r) = T(r)-T(r_0) = 0$, with $r_0$ the lowest position in data set~1 i.e. at the bottom of the slit), we can write
\begin{equation} \label{eq delta v2 with delta T = 0}
   \Delta (v^2 (r)) = v^2(r) - v^2(r_0) = \xi^2(r) - \xi^2(r_0).
\end{equation}
The flux conservation of the Alfvén waves yields 
\begin{equation}   \label{eq ksi fn rho}
   \frac{\xi(r)}{\xi(r_0)} = \left( \frac{\rho(r)}{\rho(r_0)} \right)^{-1/4}.
\end{equation}
From Eq.~\ref{eq delta v2 with delta T = 0} anf \ref{eq ksi fn rho}, we get 
\begin{equation}    \label{eq ksi0}
   \xi(r_0) = \left[ \frac{\Delta (v^2 (r))}{ \left( \frac{\rho(r)}{\rho(r_0)} \right)^{-1/2} - 1}  \right]^{1/2}.
\end{equation}
Thus, $\xi(r_0)$ can be determined for any $r$ where the hypothesis $\Delta T(r) = 0$ is true and where $\rho(r)$ and $v(r)$ can be measured. 

\begin{figure}
	\centering
		\includegraphics[width=1.00\linewidth]{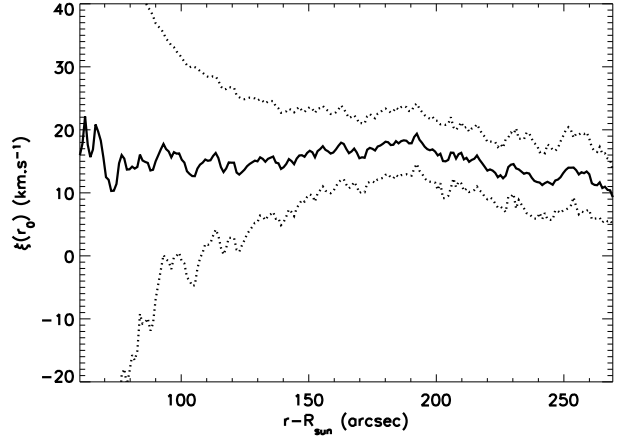}
	\caption{Value of the non-thermal velocity $\xi(r_0)$ calculated at the lowest 
	altitude observed in 
	data set~1 (i.e. $h_0 = 58\arcsec{}$ above the limb) 
	for increasing altitude $h = r - R\sun{}$ 
	(see Eq.~\ref{eq ksi0}). We used the difference in width between $r$ and $r_0$, 
	for the \ion{Mg}{x} 625~\AA{} line. 
	The dotted lines denote the error bars.}
	\label{fig xi(r_0)}
\end{figure}

Figure~\ref{fig xi(r_0)} shows that the value of $\xi(r_0)$ is relatively constant for $r$ in the range [70; 150] arcsec, which tends to prove that the $\Delta T = 0$ hypothesis is valid on that interval. 
As the  error on $\xi(r_0)$ increases when the values of $v^2(r)$ and $\rho(r)$ are close to that at $r_0$, the lowest altitudes are associated with particularly large error bars. To get a higher precision, 
we made a weighted average of $\xi(r_0)$ over the [90; 150] interval, reduced not to unnecessarily  broaden the error bars. We got $\xi(r_0) = 15 \pm 2~\textrm{km} \, \textrm{s}^{-1}$. 
%
\subsection{Determination of $\xi$ and T at every altitude} 
%
\begin{figure}
	\centering
		\includegraphics[width=1.00\linewidth]{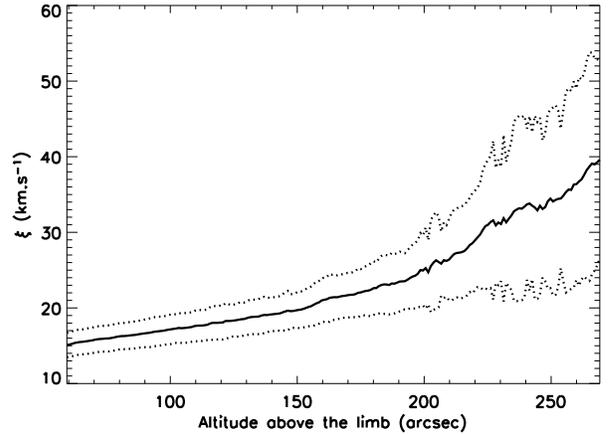}
	\caption{Non-thermal velocity $\xi$ with increasing altitude above the North polar coronal hole (data set~1), derived by using the flux conservation of Alfvén waves (Eq.~\ref{eq ksi fn rho}). The dotted lines denote the error bars. }
	\label{fig xi(r)}
\end{figure}
Once we have $\xi(r_0)$, $\xi(r)$ comes from Eq.~\ref{eq ksi fn rho} (Fig.~\ref{fig xi(r)}), every where we know $\rho(r)$. The \ion{Mg}{x} 625~\AA{} line width and the $\Delta T = 0$ hypothesis were only used to determine the constant $\xi(r_0)$. The determination of $\xi(r)$ now only requires the density diagnostic. Nevertheless, Eq.~\ref{eq ksi fn rho} is only valid if we assume that Alfvén waves undergo no damping between $r_0$ and $r$ and, of course, if Alfvén waves are entirely responsible for the appearance of the non-thermal velocity. 

Above 150\arcsec{}, the slope of the density curve changes. 
This may be due to the presence of a different structure on the plane of the sky, e.g. a difference between plume and interplume. 
If so, one would expect that the flux conservation is no longer valid above 150\arcsec{}, as the necessary continuity of the flux tubes is compromised. We nevertheless derived $\xi$ up to 270\arcsec{}, keeping in mind that the results are less reliable above 150\arcsec{}. 

\begin{figure*}
	\centering
		\includegraphics[width=0.45\linewidth]{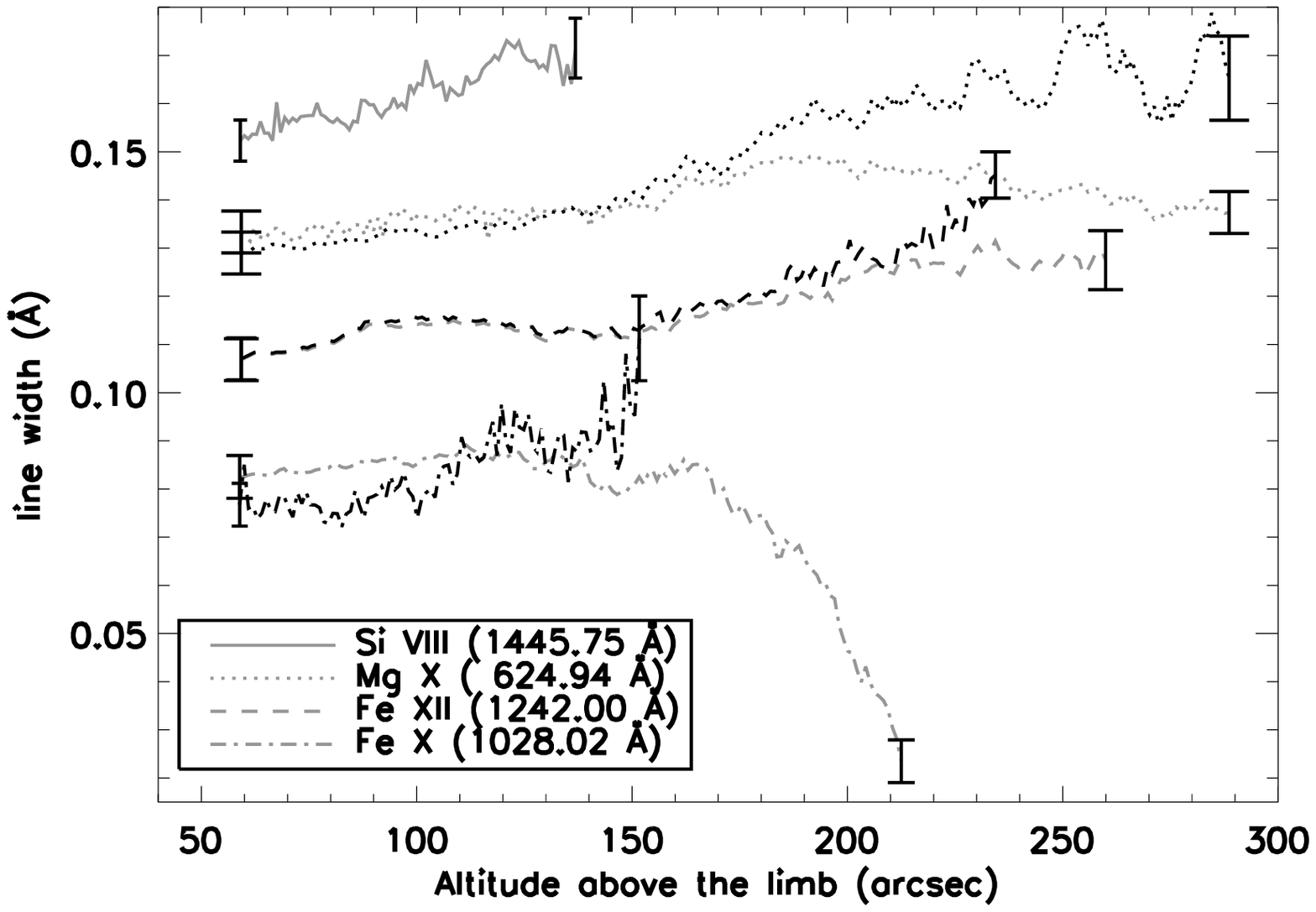}
		\includegraphics[width=0.45\linewidth]{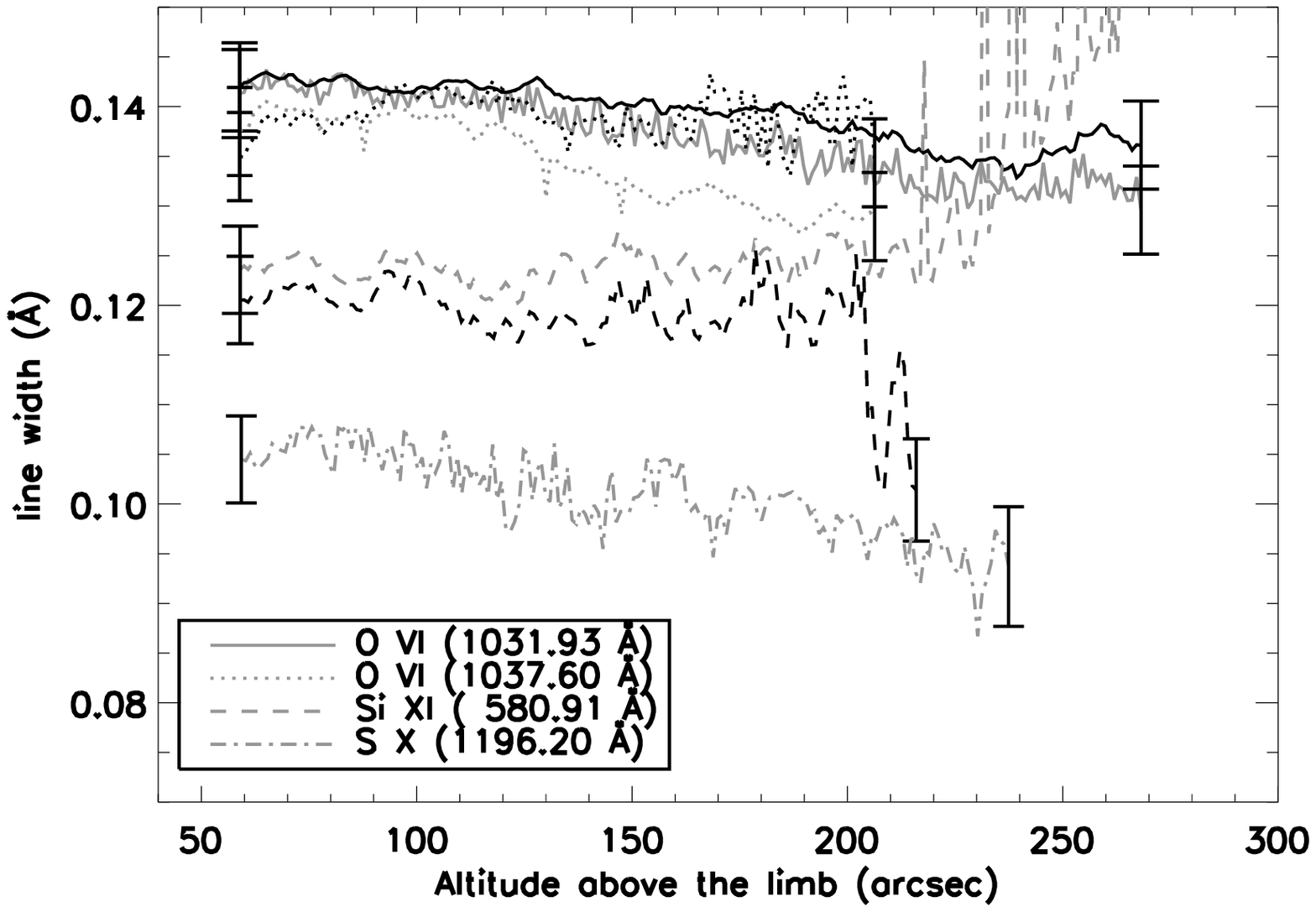}
	\caption{Line-width variation above the limb for some ions in data set~1. Spectra are created by a running sum over 40~spatial pixels. In grey, the value when the stray light is not corrected; the value after correction (in black) is not available for every ion (e.g. \ion{Si}{viii}). The appearance of a decrease when some particular altitude is reached is explained by the stray light contamination (first panel, e.g. \ion{Fe}{x} blended with \ion{O}{i}, or \ion{Mg}{x}, blended with the same line as stray light). The second panel shows some lines that start to decrease (or stay constant) from the lowest altitude that we observed (with or without correction), which apparently contradicts the behaviour of the other lines. For better visibility, the errors bars are only shown on both extremities of the curves, and the width of \ion{Fe}{x} is shifted by -0.03~\AA{} (first panel), while that of \ion{S}{x} is shifted by -0.02~\AA{} (second panel). }
	\label{fig radial sigma}
\end{figure*}

Temperature for each ion species can now be determined from the observed linewiths (shown in Fig.~\ref{fig radial sigma}) by using Eq.~\ref{eq sigma}. The altitude range is nonetheless restricted to positions where the non-thermal velocity is determined, i.e. below about 270\arcsec{}. 
To produce each spectrum, we performed a running sum over 40 spatial pixels. 
The results are shown in Fig.~\ref{fig radial T} and discussed below. 
%
\section{Results and discussion} \label{sec results}
%
\subsection{No evidence of Alfvén wave damping} 
\label{sec No evidence for damping of Alfven waves}
Before analysing more deeply the coronal line widths in terms of temperature and non-thermal velocity, we will discuss the radial variation of the width as a whole. 
This is an important issue because, in past studies, the fact that the widths of coronal lines were decreasing above some given altitude was sometimes interpreted as the signature of the damping of Alfvén waves. In fact, it appears that this decrease is essentially an effect of the instrumental stray light. 

Figure~\ref{fig radial sigma} shows some examples of the width of different lines, with increasing altitude above the limb. The \ion{Mg}{x} 625~\AA{} and \ion{Si}{viii} 1445~\AA{} lines, in particular, have been widely analysed in the literature. When no stray light correction is made, the \ion{Mg}{x} 625~\AA{} line presents the same behaviour as observed by previous authors: first, the line width increases, then reaches a maximum around 200\arcsec{} ($\approx 1.2$~R\sun{} from disc centre), and starts to decrease. We find the same kind of behaviour for the \ion{Fe}{x} 1028~\AA{} or \ion{Fe}{xii} 1242~\AA{} lines. There are not enough statistics for the \ion{Si}{viii} 1445~\AA{} line to extend the curve above 140\arcsec{}. When we correct spectra from the stray light contribution in the case of the \ion{Mg}{x} line, it is no longer possible to see any decrease, but maybe a flattening trend. For the \ion{Fe}{x} and \ion{Fe}{xii} lines, the stray light becomes so strong that there are not enough statistics to follow them as high as in the case with no correction; nevertheless, it appears an increasing trend for both lines. 

That the stray light contamination is responsible for reducing the coronal line widths observed above the limb is well known, but appears to be underestimated in most studies with SUMER, as we find that the effect is already noticeable well below 200\arcsec{}. One should question in particular any study that does not directly predicts the stray-light spectra. 

The explanation of this effect is rather simple: lines are always narrower in stray-light spectra than in off-limb spectra; therefore, when the stray light contamination grows as the altitude increases, the observed line width is reduced. It is worthwhile recalling that a stray-light spectrum is similar to any typical spectrum that could be observed on the disc in the same wavelength range. Therefore two cases are possible: either the stray-light blending is caused by the same coronal or transition region line, when it is particularly prominent on the disc (e.g. for \ion{Mg}{x} 625~\AA{} or \ion{Fe}{xii} 1242~\AA{} and 1349~\AA{}), or it is due to another line of neighbouring wavelength (e.g. \ion{Fe}{x} lines that are blended with \ion{O}{i} near 1028~\AA{} or with \ion{C}{i} near 1463.5~\AA). In this case, the blending line is usually a colder and narrower one, dominant on the disc, while the hotter one is dominant above the limb. 

When no stray light correction is made, the observation of either a plateau or a decrease after the initial increase with the altitude, and the altitude where this is observed, is then simply determined by the way the ratio of stray light to coronal emission evolves with the altitude. 
In general, lines having a lower formation  temperature than the others present a radiance that decreases more rapidly with the altitude, and are consequently subject to more contamination. 
Thus, in Fig.~\ref{fig radial sigma}, the uncorrected \ion{Fe}{x} line width starts to decrease at a lower altitude ($\approx 150\arcsec{}$) than that of \ion{Mg}{x} ($\approx 190\arcsec{}$), which also starts to decrease  at a lower altitude than that of \ion{Fe}{xii} ($\approx 230\arcsec{}$; cf. Fig.~\ref{fig T fn Tf_data set 1} for the formation temperatures of these ions). 
But from one data set to another, the same line can also present different behaviours. One can notice in Fig. ~\ref{fig Mg X different data sets} (upper panel) that the \ion{Mg}{x} 625~\AA{} line width presents a plateau from 60\arcsec{} to 180\arcsec{}, while it reaches a maximum at three different altitudes in data sets 1, 3, and 4. 
For these sets, the difference in altitude of the maximum present some correlation with the observed radiance (bottom panel). 
Note that if the observed radiance consists of the addition of the intrinsic and of the stray light radiance, the last one is negligible at low altitude, so that the curves give a good idea of the intrinsic radiance there. 
Thus, the curve starts to decrease before all the others in set~3 ($\approx 150\arcsec$) and reaches a plateau in set~4 (still around 150\arcsec) before the maximum seen in set~1 ($\approx 230\arcsec$). This fact can be explained by an almost identical stray light contribution in all data sets, dominating more rapidly in data sets where the intrinsic emission is smaller. 

There is, however, one reservation concerning the interpretation of the decrease in the line width in terms of stray light contamination: even after correction, there is still a decrease, or an almost constant value, for some lines in data set~1: both \ion{O}{vi}, the \ion{Si}{xi} 580.91~\AA{}, and the \ion{S}{x} 1196.20~\AA{} lines (Fig.~\ref{fig radial sigma}, right panel). For these lines, there is no increase phase, as this decreasing trend begins at 57\arcsec{}. Once the non-thermal velocity is removed, the derived temperatures decrease with increasing altitude (Fig.~\ref{fig radial T}, third panel). We have no explanation for this behaviour, as these lines are not intense enough on-disc to be subject to photoexcitation, except for \ion{O}{vi}. It is possible that these lines are affected by some blends. 
%
\subsection{Signatures of preferential heating} \label{sec Signatures of preferential heating}
%
\begin{figure}
	\centering
		\includegraphics[height=0.25\textheight]{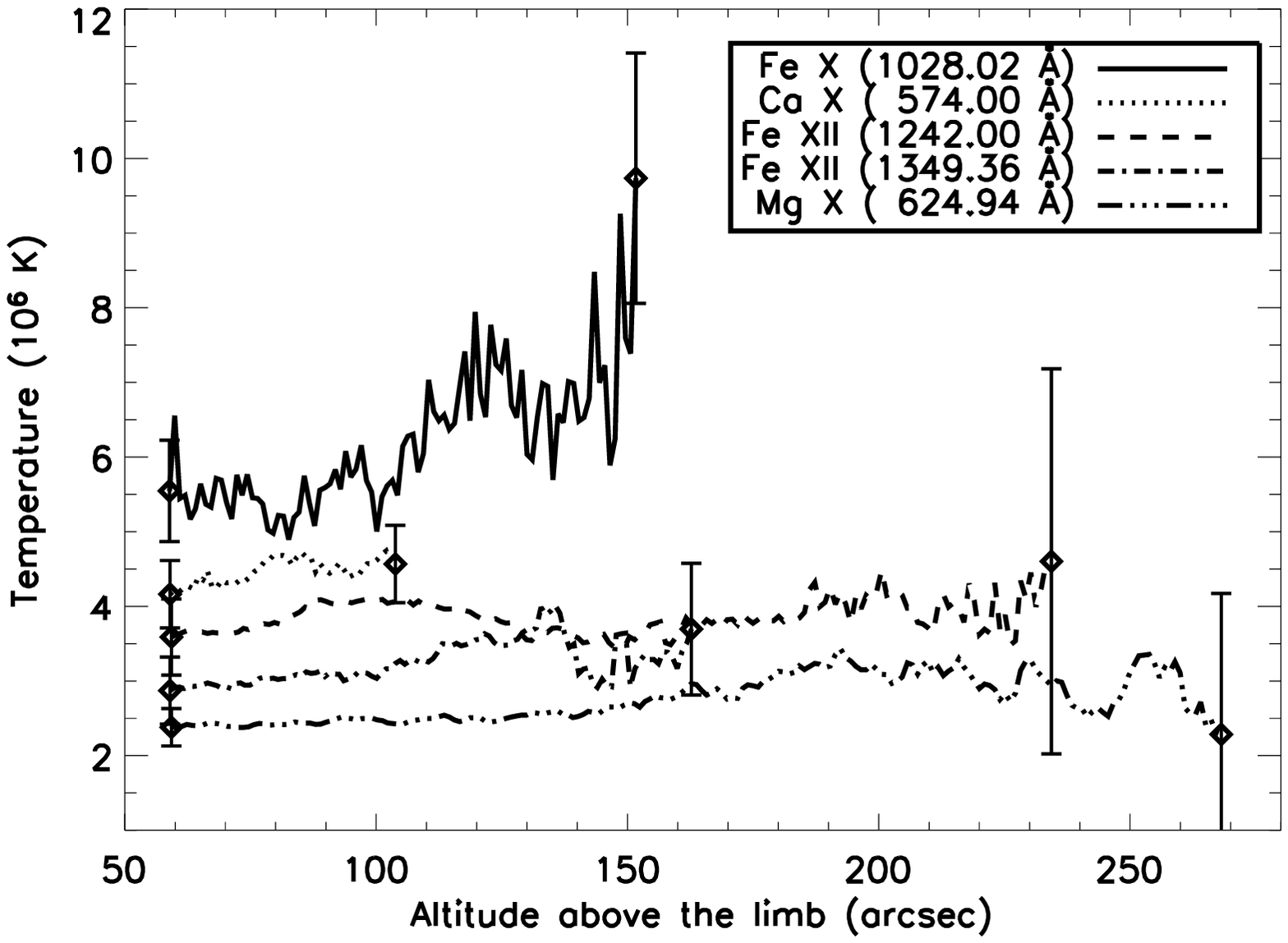}
		\includegraphics[height=0.25\textheight]{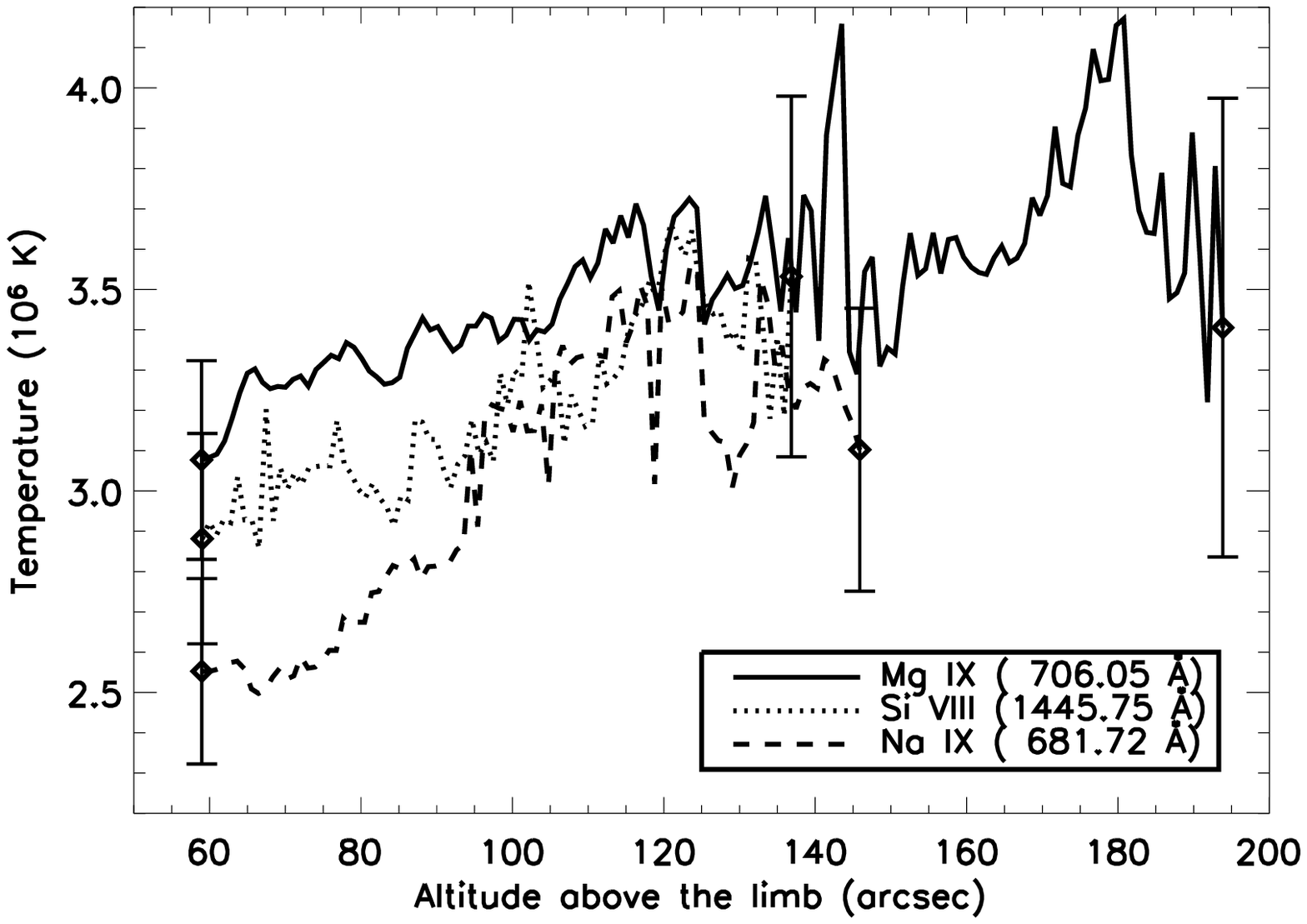}
		\includegraphics[height=0.25\textheight]{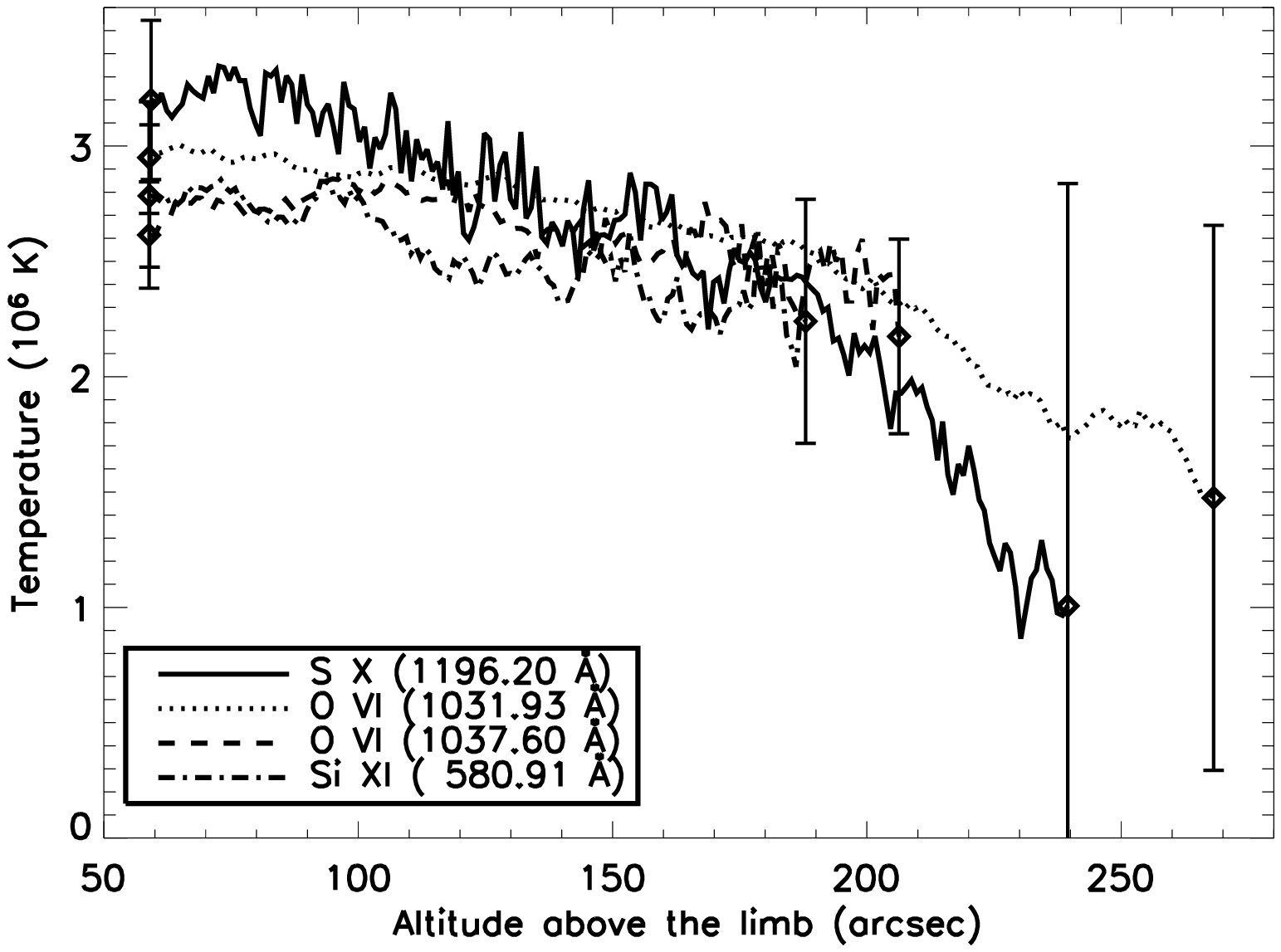}
	\caption{Temperature with increasing altitude above the limb (data set~1; spectra are produced by a running sum over 40~spatial pixels). }
	\label{fig radial T}
\end{figure}
\begin{figure}
	\centering
		\includegraphics[width=\linewidth]{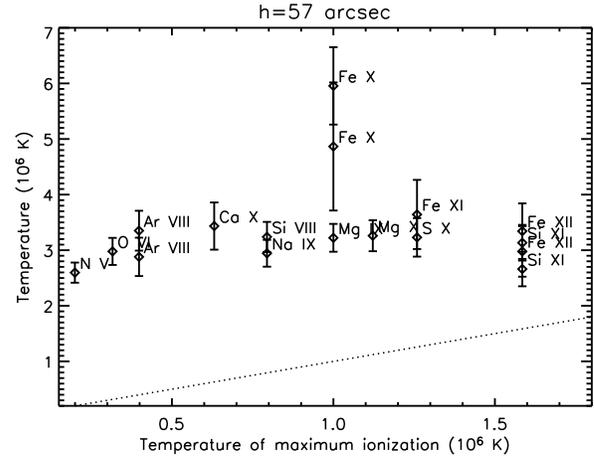}
	\caption{Measured ion temperatures (data set~1, 57\arcsec{} above the limb) as functions of the maximum ionization temperatures \citep[taken from][]{Arnaud85}. They are well above the corresponding ionization temperature (dotted line). }
	\label{fig T fn Tf_data set 1}
\end{figure}

The ion temperatures in Fig.~\ref{fig radial T} are of the same order of magnitude as what is found in previous studies \citep[e.g.][]{Tu98, Patsourakos02}; they all exceed $2 \times 10^6~\textrm{K}$  at 57\arcsec{}. They exhibit a wide range of values, which makes the isothermal hypothesis unlikely. The temperature of the three iron lines, for example,  are arranged in the opposite order to their corresponding ionization temperature (Fig.~\ref{fig T fn Tf_data set 1}). Then the LOS effect associated to the stratification in temperature of the corona also appears unlikely. 
The temperature increases with the altitude for most of the lines, except in the bottom frame  of  Fig.~\ref{fig radial T} corresponding to the exceptions mentioned above. This implies some heating mechanism. The relative discrepancy between both temperatures yielded by the \ion{Fe}{xii} lines at 1242 and 1349~\AA{}, below 130\arcsec{}, may be explained by the blending of the 1242~\AA{} line with an \ion{Si}{x} line \citep[cf. Table~1 in][]{Feldman97}, as well as by the error bars. The temperature of \ion{Fe}{x} is very high, but it is confirmed at low altitude by a second \ion{Fe}{x} line, at 1463~\AA{} (see Fig.~\ref{fig T fn q/m_data set 1}). 

\begin{figure*}
	\centering		   
	\includegraphics[width=0.45\linewidth]{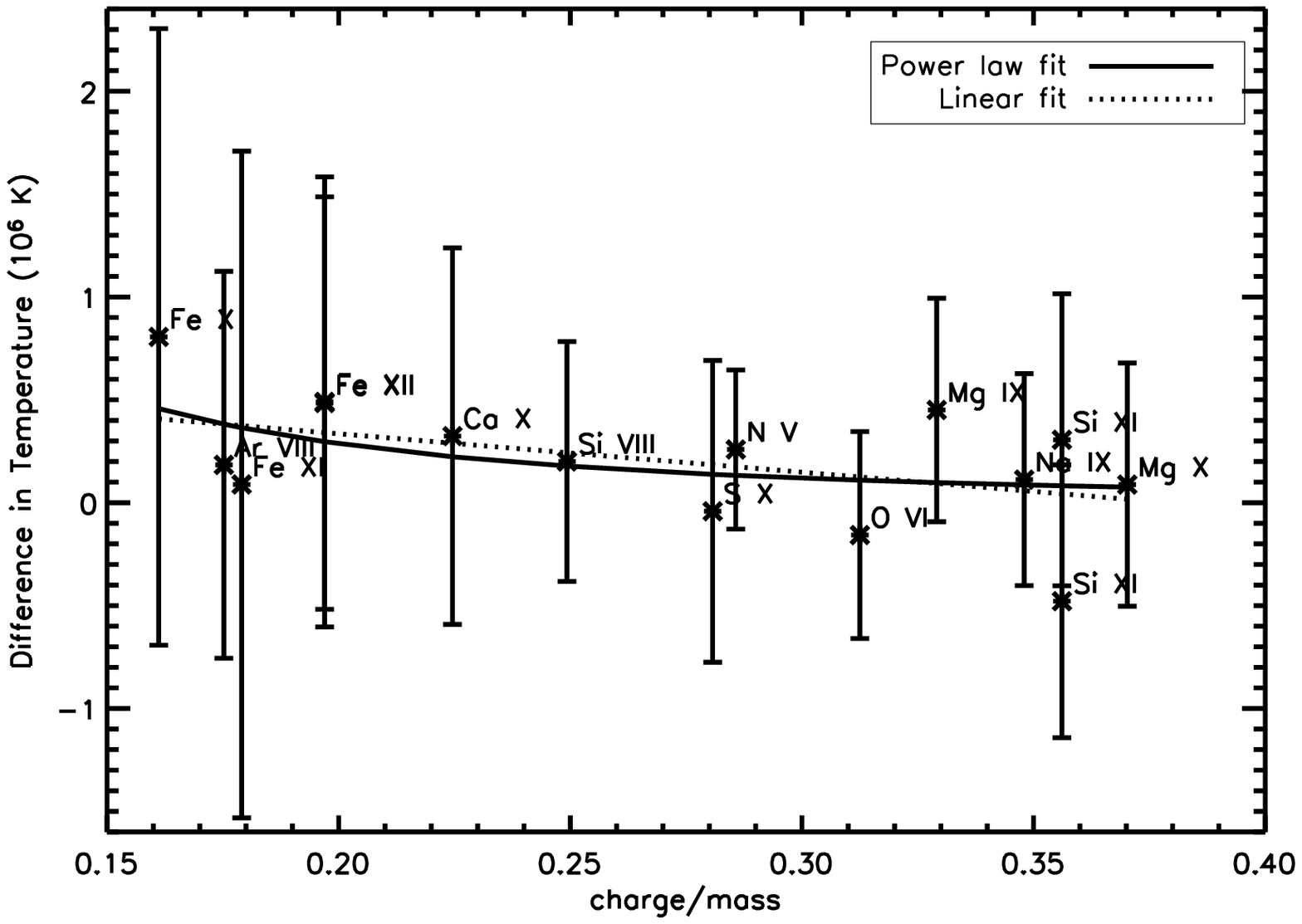}
  \includegraphics[width=0.45\linewidth]{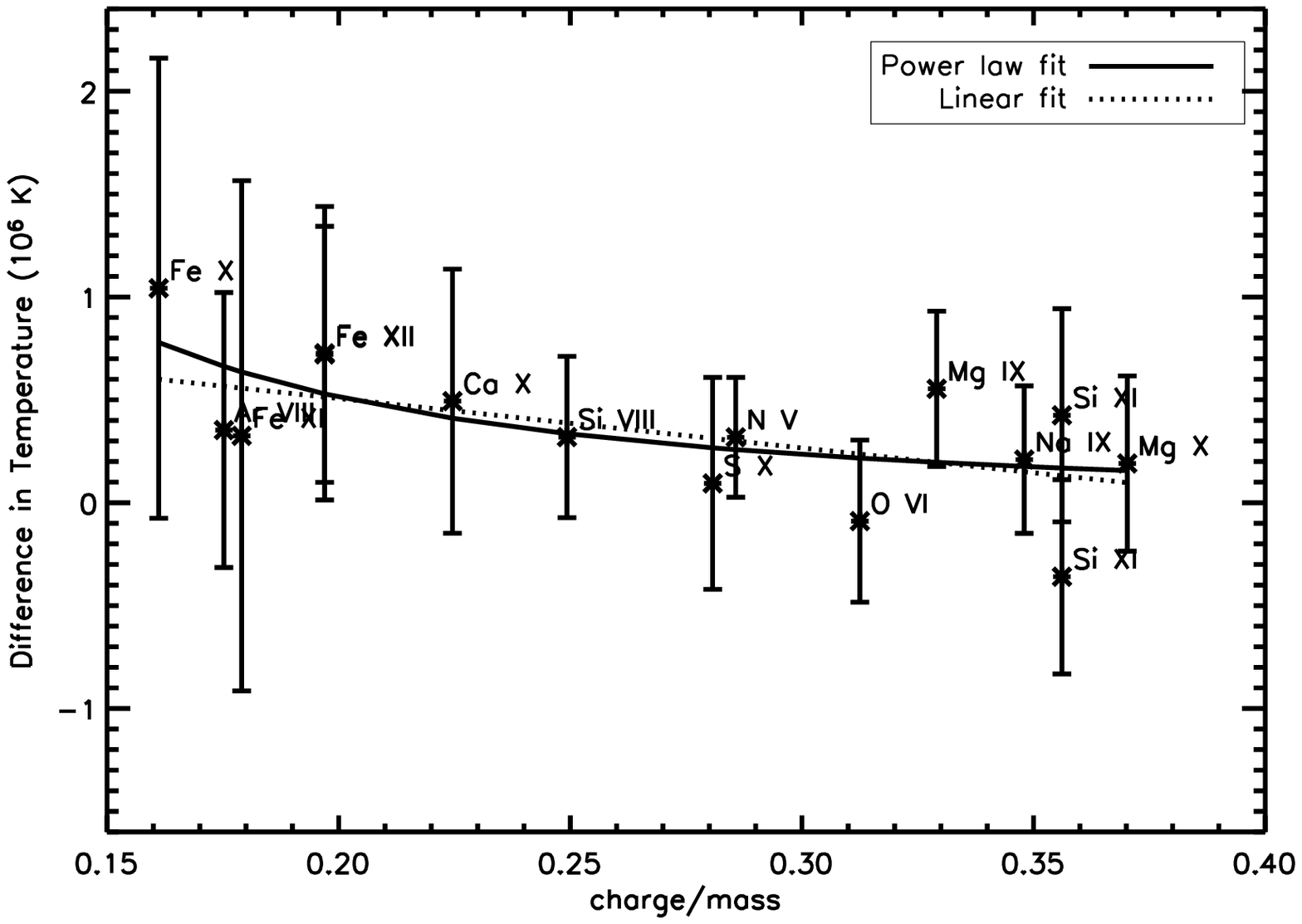}
	\caption{Difference in temperature between 57\arcsec{} and 102\arcsec{}, once the non-thermal contribution is removed (left panel). The right panel corresponds to the case where we suppose that there is no variation in non-thermal velocity, only a temperature one. This case is equivalent to the case $\xi = 0$, i.e. when no non-thermal velocity is considered at all. 
	The fitting functions are discussed in Sect.~\ref{sec Signatures of preferential heating}. The slopes of the linear fits are $-1.9 \pm 3.1$ and $-2.4 \pm 2.1$, respectively, in the left and the right panels (in $10^6~\textrm{K}$ per unit of normalized $q/m$). }
	\label{fig diff T}
\end{figure*}

\begin{figure*}
	\centering
		\includegraphics[width=0.45\linewidth]{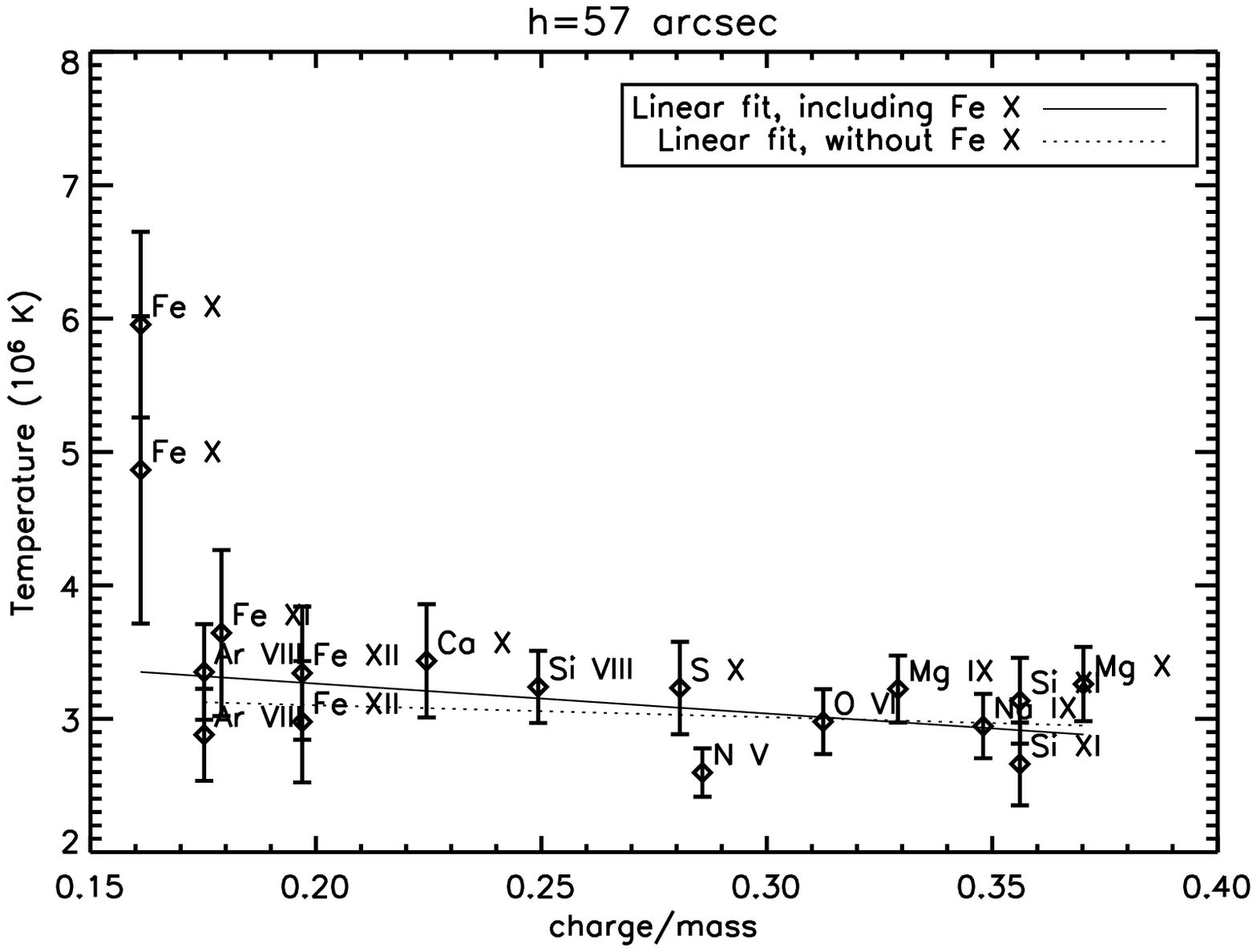}
		\includegraphics[width=0.45\linewidth]{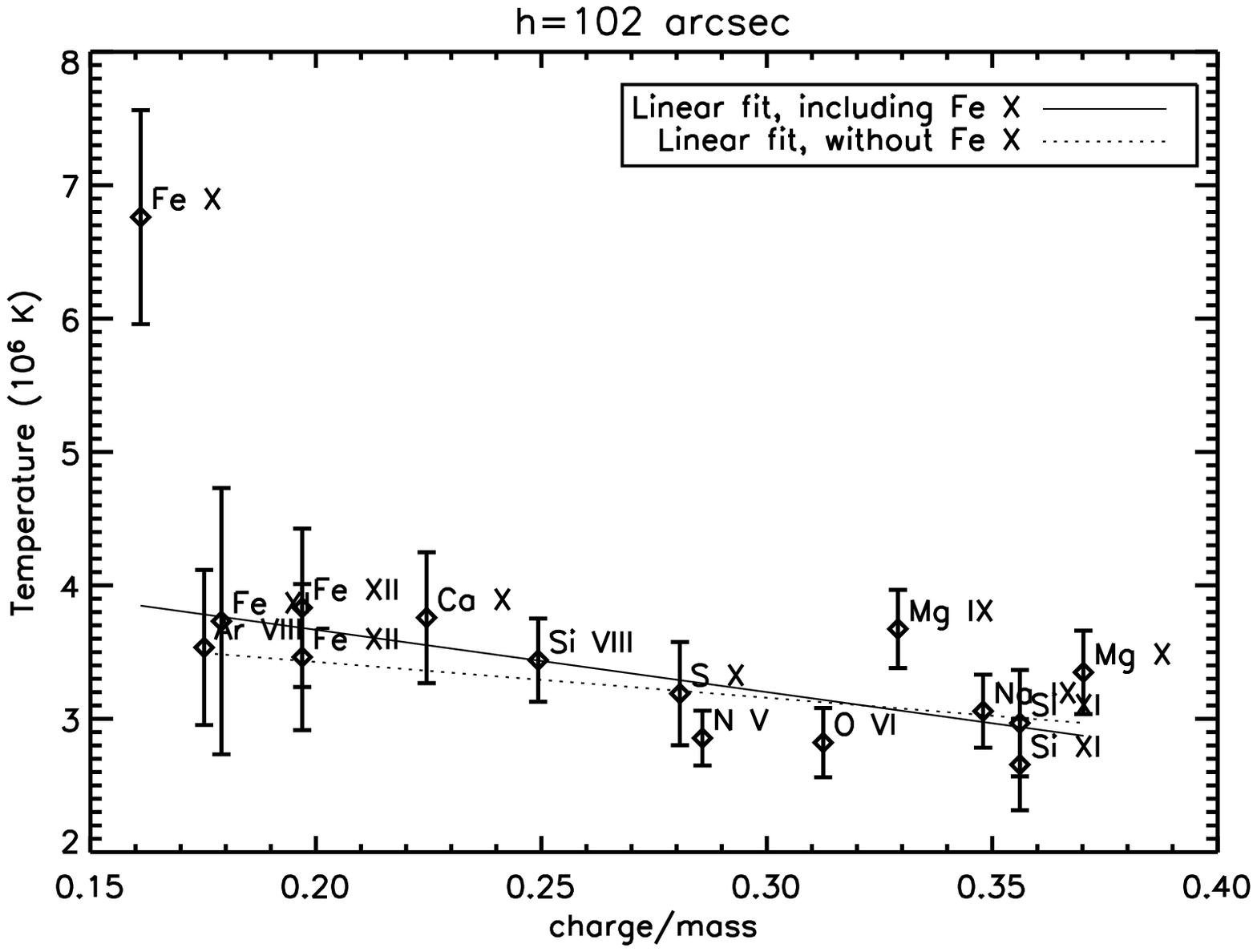}
	\caption{Temperatures as functions of the charge-to-mass ratio $q/m$, 57\arcsec{} (left panel) and 102\arcsec{} (right panel) above a polar coronal hole (data set~1; the non-thermal velocity has been removed from the line widths). The interest of making a linear fit is discussed in Sect.~\ref{sec Signatures of preferential heating}. }
	\label{fig T fn q/m_data set 1}
\end{figure*}

To reveal a preferential heating, we plot in Fig.~\ref{fig diff T} the difference in temperature as a function of the $q/m$ ratio between 57\arcsec{} and 102\arcsec{}, for as many lines as possible in data set~1. The line widths and temperatures measured at these altitudes are presented in Table~\ref{tab results data set 1}. 
It is important to note that, for lines seen in second order, the widths are twice as large as what would be measured in first order. This is taken into account when using Eq.~\ref{eq sigma} and the derived ones. 
We average over 30~pixels at 57\arcsec{} and 60~pixels at 102\arcsec{}, to exploit some very faint lines. There were not enough statistics at 102\arcsec{} to analyse the \ion{Fe}{x} line at 1463~\AA{}, so that we only keep the one at 1028~\AA{}. In the left panel, the non-thermal velocity has been taken into account, using results in Fig.~\ref{fig xi(r)}: 15 and $17~\textrm{km} \, \textrm{s}^{-1}$ at 57\arcsec{} and 102\arcsec{}, respectively. The right panel corresponds to the case where no variation in non-thermal velocity is considered between 57\arcsec{} and 102\arcsec{}, which is discussed in Sect.~\ref{sec Derived non-thermal velocity and its effect}. 
The large error bars are due to the errors propagated from both the observed and the stray-light spectra to the line width (Gaussian-fit procedure) and the error on the determination of the non-thermal velocity. 
The weighted power law and linear fits show a trend for the ion species having the lowest $q/m$ ratios to experience more heating than the others, which is compatible with ion-cyclotron preferential heating. Both kinds of fit curves are not very different from each other. We do not intend to fit the data with any particular theoretical model. However, the power law tends to 0 for high $q/m$, while the linear fit, with a slope of $-1.9 \pm 3.1$ (in $10^6~\textrm{K}$ per unit of normalized $q/m$), plummets towards $-10^6~\textrm{K}$ for $q/m = 1$ (protons). Therefore, a power law seems to better represent a preferential heating of low $q/m$ species, although the correct function is probably much more complicated. 
One can notice that the different line width behaviours observed by \citet{Singh03,Singh03b}, which they link to the formation temperature (LOS effect), are also compatible with ion-cyclotron heating. It is worthwhile noting that they correct for the instrumental stray light  with the same method as we do. The case of \ion{Fe}{xiv}, in their study, the line width of which  decreases with the altitude, is nevertheless disturbing. But this is somehow similar to the behaviour of some lines in our data set~1 (Fig.~\ref{fig radial T}, bottom panel). If it is a real variation in temperature, this apparent "cooling" process should be explained. We notice, though, that all these lines are associated to the highest $q/m$, meaning that they are less likely to be concerned by ion-cyclotron heating. Therefore, it is possible that all other ions experience the same kind of cooling,  or a decrease in non-thermal velocity, but that this effect is counterbalanced by the preferential heating for the ions having the lowest $q/m$. 

A second kind of analysis consists in plotting the ion temperatures as functions of $q/m$, at a given altitude (Fig.~\ref{fig T fn q/m_data set 1}). It shows that the ion species having the lowest $q/m$ have a higher temperature than the others. The trend is emphasised more at 102\arcsec{} than at 57\arcsec{}, which is consistent with the previous analysis. The particularly high temperature of \ion{Fe}{x} tends to compress the graph, but the trend does not rely on this ion alone. To show that, we use a weighted linear fit including the \ion{Fe}{x} data or not. Even if a straight line is obviously not the proper function to fit the data, it makes the comparison simpler, while no theoretical function is available. At 57\arcsec{}, the slope is equal to $-2.2 \pm 1.2$ with \ion{Fe}{x} and $-0.9 \pm 1.3$ without it, still in $10^6~\textrm{K}$ per unit of normalized $q/m$. At 102\arcsec{}, it gives $-4.7 \pm 1.7$ with \ion{Fe}{x} and $-2.7 \pm 1.8$ without it. 

To interpret these curves, some points must be emphasised. First, the protons can be considered as a cooling source for all minor ion species that could be subject to any preferential heating. They compose the largest part of the coronal plasma so that collisions of one given minor ion are much more frequent with protons than with other species. As the $q/m$ ratio of the protons, equal to 1 with our normalization, is higher than those of all the minor ions, they are also least likely to experience cyclotron heating. Consequently, their temperature must be lower than or equal to that of the ions having the highest $q/m$ ratios, and the temperature of all other species is the result of a competition between preferential heating and cooling with the protons. By measuring differences in temperature, one only has access to the net heating. 
Second, most of the ion-cyclotron numerical simulations show that the species having the lowest $q/m$ absorb most of the wave power available, to the detriment of the others that cannot resonate efficiently yet \citep[e.g.][]{Cranmer00, Vocks02b}. This could explain the high temperature of \ion{Fe}{x} as compared to the other species. 

That low $q/m$ ions are already hotter than the others at 57\arcsec{} suggests that ion-cyclotron resonance has already taken place at lower altitudes. 
Such a preferential heating is still at work between 57\arcsec{} and 102\arcsec{} (cf. Fig.~\ref{fig diff T}). 
If the net heating experienced by the lowest $q/m$ ion species, as well as their temperatures, appear high, one should keep in mind, though, that these are minor ions. Therefore, this heating does not require much energy, nothing comparable to what is needed to appreciably heat the protons. 
And yet, even if we have no direct measurement of the proton temperature here, one can speculate that it is not very different from those of the ions species having the higher $q/m$ ratios. 
One can firmly suppose from the discussion in the previous paragraph, indeed, that the potential cyclotron heating for the high $q/m$ species is low. Then, it is not likely that large temperature differences with the protons can be maintained against collisions. 
If the proton temperature is as high as those of the colder ion species observed, this is far more than the electron temperature usually observed in coronal holes \citep[$\approx 8 \times 10^5$~K,][]{habbal93, Wilhelm98, David98}. 
Then, one has to consider that the protons may have been heated in comparison with the electrons, again at altitudes below 57\arcsec{}. Investigating the helium or hydrogen temperatures and their variations with the altitude would be a big help for getting more information in the $q/m$ analysis, but the associated lines in the SUMER spectrum suffer from a strong stray-light blending. 

\begin{table*}
 \caption{Line widths measured in data set~1 (coronal hole) at two positions above the solar limb, and the associated ion temperatures. } 
 \label{tab results data set 1}
 \centering
    \begin{tabular}{l c c c c c  c c c}
       \hline
       \hline
           &      
           & \multicolumn{3}{c}{57\arcsec{} (n$_\textrm{e}=4.8 \times 10^7~\textrm{cm}^{-3}, 
             \xi=15~\textrm{km} \, \textrm{s}^{-1}$)} 
           & \multicolumn{3}{c}{102\arcsec{} (n$_\textrm{e}=2.7 \times 10^7~\textrm{cm}^{-3}, 
             \xi=17~\textrm{km} \, \textrm{s}^{-1}$)} \\
       Line$^{\mathrm{a}}$ & $q/m^{\mathrm{b}}$ 
          & Observed width$^{\mathrm{c}}$ &  Width$^{\mathrm{d}}$  & T$_i^{\mathrm{e}}$ 
          & Observed width$^{\mathrm{c}}$ &  Width$^{\mathrm{d}}$  & T$_i^{\mathrm{e}}$ \\
       {} & {} & (\AA) & (\AA) & ($10^6$~K) & (\AA) & (\AA) & ($10^6$~K) \\ 
       \hline
       \ion{Fe}{x}   1028.02 &  0.16 
         	&  0.1111 $\pm$ 0.004  &  0.1102 $\pm$ 0.004 &  5.96 $\pm$ 0.70 
         	&  0.1161 $\pm$ 0.004  &  0.1184 $\pm$ 0.004  &  6.76 $\pm$ 0.80  \\
       \ion{Fe}{x}   1463.50  &  0.16 
         	&  0.1434 $\pm$ 0.004  &  0.1435 $\pm$ 0.013  &  4.87 $\pm$ 1.15    
          &        \ldots        &        \ldots        &       \ldots      \\
       \ion{Fe}{xi}   1467.08 &  0.18 
         &  0.1310 $\pm$ 0.004  &  0.1272 $\pm$ 0.007  &  3.64 $\pm$ 0.62    
         &  0.1376 $\pm$ 0.004  &  0.1318 $\pm$ 0.011  &  3.73 $\pm$ 1.00  \\
       \ion{Ar}{viii}   700.26$^\ast$  &  0.18 
         &  0.1158 $\pm$ 0.004  &  0.1351 $\pm$ 0.002  &  3.35 $\pm$ 0.36                                &  0.1222 $\pm$ 0.005  &  0.1411 $\pm$ 0.004  &  3.53 $\pm$ 0.58  \\
       \ion{Ar}{viii}    713.82$^\ast$  &  0.18 
         &  0.1291 $\pm$ 0.004  &           \ldots        &  2.88 $\pm$ 0.34    
         &        \ldots        &        \ldots        &       \ldots      \\
       \ion{Fe}{xii}   1242.00 &   0.20 
         &  0.1035 $\pm$ 0.004  &  0.1039 $\pm$ 0.004  &  3.34 $\pm$ 0.50    
         &  0.1132 $\pm$ 0.004  &  0.1128 $\pm$ 0.004  &  3.83 $\pm$ 0.59  \\
       \ion{Fe}{xii}   1349.36  & 0.20 
          &  0.1071 $\pm$ 0.004  &  0.1078 $\pm$ 0.004  &  2.98 $\pm$ 0.45    
          &  0.1176 $\pm$ 0.004  &  0.1177 $\pm$ 0.004  &  3.46 $\pm$ 0.55  \\
       \ion{Ca}{x}    574.00$^\ast$  & 0.22 
          &  0.1111 $\pm$ 0.004  &  0.1118 $\pm$ 0.002  &  3.43 $\pm$ 0.42    
          &  0.1137 $\pm$ 0.004  &  0.1185 $\pm$ 0.002  &  3.76 $\pm$ 0.49  \\
       \ion{Si}{viii}   1445.75  & 0.25 
          &  0.1605 $\pm$ 0.004  &         \ldots         &  3.24 $\pm$ 0.27    
          &  0.1674 $\pm$ 0.004  &         \ldots       &  3.44 $\pm$ 0.31  \\
       \ion{S}{x}   1196.20  & 0.28 
          &  0.1250 $\pm$ 0.004  &          \ldots       &  3.23 $\pm$ 0.35    
          &  0.1266 $\pm$ 0.004  &           \ldots      &  3.19 $\pm$ 0.39  \\
       \ion{N}{v}   1238.81  & 0.29 
          &  0.1532 $\pm$ 0.004  &  0.1709 $\pm$ 0.004  &  2.60 $\pm$ 0.18    
          &  0.1448 $\pm$ 0.004  &  0.1804 $\pm$ 0.004  &  2.86 $\pm$ 0.21  \\
       \ion{O}{vi}   1031.93  & 0.31 
          &  0.1426 $\pm$ 0.004  &          \ldots        &  2.98 $\pm$ 0.24    
          &  0.1406 $\pm$ 0.004  &          \ldots      &  2.82 $\pm$ 0.26  \\
       \ion{Mg}{ix}    706.05$^\ast$  & 0.33 
          &  0.1669 $\pm$ 0.004  &  0.1669 $\pm$ 0.002  &  3.22 $\pm$ 0.25    
          &  0.1775 $\pm$ 0.004  &  0.1795 $\pm$ 0.002  &  3.67 $\pm$ 0.29  \\
       \ion{Na}{ix}    681.72$^\ast$  & 0.35 
          &  0.1561 $\pm$ 0.004  &  0.1587 $\pm$ 0.002  &  2.94 $\pm$ 0.24    
          &  0.1535 $\pm$ 0.004  &  0.1637 $\pm$ 0.002  &  3.06 $\pm$ 0.27  \\
       \ion{Si}{xi}    580.91$^\ast$   & 0.36 
          &  0.1269 $\pm$ 0.004  &  0.1271 $\pm$ 0.002  &  3.13 $\pm$ 0.32    
          &  0.1251 $\pm$ 0.004  &  0.1204 $\pm$ 0.002  &  2.66 $\pm$ 0.34  \\
       \ion{Si}{xi}    604.15$^{\ast, \mathrm{f}}$   & 0.36 
          &  0.1229 $\pm$ 0.005  &          \ldots        &  2.66 $\pm$ 0.31    
          &  0.1313 $\pm$ 0.005  &          \ldots         &  2.97 $\pm$ 0.40  \\
       \ion{Mg}{x}    624.94$^\ast$   & 0.37 
          &  0.1477 $\pm$ 0.004  &  0.1485 $\pm$ 0.002  &  3.26 $\pm$ 0.28    
          &  0.1513 $\pm$ 0.004  &  0.1524 $\pm$ 0.002  &  3.35 $\pm$ 0.31  \\
       \hline
    \end{tabular}
  \begin{list}{}{}
  \item[$^{\mathrm{a}}$] Wavelengths in \aa ngströms are taken from \citet{Feldman97}. Lines marked with an asterisk are observed in second order. 
  \item[$^{\mathrm{b}}$] Ion charge-to-mass ratio. 
  \item[$^{\mathrm{c}}$] Half-width at 1/$\sqrt{e}$ (Gaussian). Instrumental width is removed, but spectra are not corrected from the instrumental stray light. For lines seen in second order, the widths are twice as large as what would be measured in first order (see Sect.~\ref{sec Signatures of preferential heating}). 
  \item[$^{\mathrm{d}}$] Half-width at 1/$\sqrt{e}$ when spectra are corrected from the stray light, with the instrumental width removed. 
  \item[$^{\mathrm{e}}$] Ion temperature deduced from the corrected line width, with the non-thermal velocity removed. When no correction was necessary or possible, the observed width was used. 
  \item[$^{\mathrm{f}}$] This line is blended with an \ion{Fe}{vii} line, as given by \citet{Ekberg03}. See comments in Sect.~\ref{sec correction from the instrumental stray light}. 
  \end{list}
\end{table*}
%
\subsection{Derived non-thermal velocity and its effect on the signature} \label{sec Derived non-thermal velocity and its effect}
%
The signatures of preferential heating are sensitive to the derived non-thermal velocity. In particular, any variation in width is distributed among variations in temperature and in non-thermal velocity: 
\begin{equation} \label{delta sigma}
   \delta \sigma = \frac{\lambda^2}{2 \sigma c^2} (\frac{k}{m}  \, \delta T 
   + \xi \, \delta \xi). 
\end{equation}
Even if there is little variation in $\xi$ between 57\arcsec{} and 102\arcsec{}  (2~$\textrm{km} \, \textrm{s}^{-1}$), the actual value of $\xi$ has a strong effect on the interpretation of the line-width variation. The value of $\xi(r)$ was derived under flux conservation condition, with no damping occurring for the Alfvén waves. We have already said that it was not possible to verify this assumption, because the \ion{Mg}{x} 625~\AA{} line width keeps increasing, so a possible decrease in $\xi$ may be hidden by an increase in temperature. If $\delta \xi(r)$ starts to decrease, in reality, above some  position $r$ or even becomes negative, i.e. the damping is greater than the amplitude increase due to the flux conservation, then the line-width growth only has to be attributed to a growth of temperature. But this situation may only amplify the preferential heating trend, both in 
absolute temperatures 
and in differences in temperature,  as shown in the right panel of Fig.~\ref{fig diff T}. Note that the case $\delta \xi = 0$ is equivalent to the case $\xi(r_0) = 0~\textrm{km} \, \textrm{s}^{-1}$ (cf. Eq~\ref{delta sigma}). 
In contrast, if $\xi(r)$ is larger than what we find, considering for example that Alfvén waves are not responsible for the apparent non-thermal velocity, hence that $\xi$ is not constrained by the flux conservation, then the preferential heating trends are weakened. 

The non-thermal velocities that we got at low altitude in the coronal hole are within the lower limit of previously published results. However, most of those results were obtained by using the formation temperature hypothesis, which underestimates the ion temperatures. 

A value of 25 or 30~$\textrm{km} \, \textrm{s}^{-1}$ for the alfvénic fluctuations is typically used as a boundary condition at the coronal base in numerical simulations of heating and acceleration of the fast solar wind (see references in Sect.~\ref{sec Ion cyclotron resonance and Alfvén waves}). However, this value is only 
encountered above 200\arcsec{} in our results. There is then a lack of energy at the coronal base to reproduce the speed and temperature of the fast solar wind by using these models. 
%
%
\section{Conclusions} \label{sec conclusion}
Alfvén waves and ion-cyclotron resonance, i.e. absorption of the high frequencies, are commonly used in theoretical models to provide the corona with heating and/or the fast solar wind with acceleration. Signatures of ion-cyclotron heating have been reported in the upper corona ($> 2.5$~R\sun) and in the solar wind. It appears important, then, to investigate ion-cyclotron resonance closer to the Sun, where the coronal line widths offer the opportunity to study both the temperature of different ion species and the amplitude of possible Alfvén waves. This is of course a challenging task as these quantities are merged into one observable, the line width. Most of the time,  methods were used that excluded \emph{a priori} the possibility of analysing both contributions simultaneously. Our analysis, on the contrary, achieves a synthesis of different approaches that can be found separately in the literature, to both test and constrain the presence of Alfvén waves and the preferential heating. We constrained the former by the gradient of the \ion{Mg}{x} 625~\AA{} line width, while the latter appears from the gradient of ion temperatures and from the distinct ion temperatures measured at a given altitude. 

Our analysis rests on the following hypotheses: i)~the coronal lines broadening is solely due to the thermal and non-thermal Doppler effects through Gaussian contributions. ii)~The non-thermal velocity $\xi$ comes from the integration of fluid motions of the coronal plasma on both the observed solid angle and the exposure time, and is identical for all ion species at a given altitude. iii)~These motions are due to the presence of Alfvén waves; if there is no damping, their amplitude increases with the altitude to conserve the energy flux in a density-stratified corona. iv)~The increase in width of the \ion{Mg}{x} 625~\AA{} line (the highest $q/m$), at low altitude, results only from undamped Alfvén waves, i.e. $\Delta T = 0$, so that a boundary constraint can be set for $\xi$ using a density diagnostic (\ion{Si}{viii} line ratio) and the energy flux conservation. v)~The Alfvén waves stay undamped on the entire observed spatial range, so that $\xi(r)$ can be derived. 

Several results must be emphasised. The decrease in the coronal line widths, occurring  around 1.2~R\sun{} for \ion{Mg}{x} for example, is essentially an effect of the instrumental stray light contamination that disappear when we remove the predicted stray-light spectrum. 
We cannot rule out, though, the existence of a plateau at higher altitudes than we observed. This result is in accordance with some theoretical ones that find it difficult to damp the Alfvén waves in the lower corona \citep{Parker91}. 

The derived non-thermal velocity is about $15~\textrm{km} \, \textrm{s}^{-1}$ at 60\arcsec{}, while a value of $25~\textrm{km} \, \textrm{s}^{-1}$ is only reached at 200\arcsec{} above the limb. This means a widely lower wave energy than that  commonly used in numerical simulations of heating and acceleration of the fast solar wind. Consequently, one has to find another form of energy, or of dissipation process, to replace or at least complement the processes usually presented in models based on alfvénic fluctuations. 

Two kinds of signatures of preferential heating are found: ion species having the lowest $q/m$ ratios are the hottest ones at a given altitude and experience more heating as the altitude increases from 57\arcsec{} to 102\arcsec{}. If some damping of the Alfvén waves occurred, i.e. if hypothesis v) was wrong, then these trends would be stronger. Yet, this is neither a proof by itself of cyclotron resonance occurring in the corona, nor does it imply that this is a dominant energy source for acceleration of the fast solar wind or the heating of the corona: the energetics have to be analysed in more depth. Minor ions can only help to trace the cyclotron resonance that may also concern the protons. 
The ion temperatures and probably the proton one are higher than the electron temperature usually measured in coronal holes, which imply some preferential heating that must have taken place below the lowest altitude where we observed, i.e. 57\arcsec{}. 

Some facts may possibly weaken our analysis. First, the width of some lines starts to decrease at the lowest altitude we observed. This contradicts our interpretation in terms of a common increase in the wave amplitude, but this can be due to another contribution in the line width,  influencing these lines more. Second, the lines associated with the lowest $q/m$ ratios  (\ion{Fe}{x} and \ion{Fe}{xi}, particularly) are highly contaminated with stray light, so that it is difficult to follow their behaviour at higher altitudes. We are also aware of the large error bars in our derived quantities. 
Therefore, our results have to be confirmed by additional observations and further analyses. 

This work also emphasises some needs for future instrumentation, especially in terms of spectrometers. The study of line widths emitted in the low corona requires a small instrumental width as compared to the intrinsic width. 
It is important to develop a better rejection of the light coming from the solar disc, for off-limb observations. 
Furthermore, a solar probe could achieve important \emph{in situ} observations to complement the remote-sensing observations, and even calibrate their interpretation.  
\begin{acknowledgements}
We wish to thank P. Lemaire, A. Gabriel, J.~C. Vial, G.~Nigro and L. Teriaca for useful discussions, and K.~Bocchialini for organising the MEDOC Campaigns at IAS (\url{http://www.medoc-ias.u-psud.fr}), which made this work possible. We also want to thanks the anonymous referee who made valuable suggestions to improve the quality of this article. The density diagnostic was made by using data from JOP 158. 
SOHO is a mission of international cooperation between ESA and
NASA. The SUMER project is financially supported by DLR, CNES, NASA, 
and ESA PRODEX Programme (Swiss contribution). 
\end{acknowledgements}
\bibliographystyle{aa} 
\bibliography{D:/LateX/biblio_database}

\begin{thebibliography}{70}
\expandafter\ifx\csname natexlab\endcsname\relax\def\natexlab#1{#1}\fi

\bibitem[{{Arnaud} \& {Rothenflug}(1985)}]{Arnaud85}
{Arnaud}, M. \& {Rothenflug}, R. 1985, \aaps, 60, 425

\bibitem[{{Banerjee} {et~al.}(2000){Banerjee}, {Teriaca}, {Doyle}, \&
  {Lemaire}}]{Banerjee00}
{Banerjee}, D., {Teriaca}, L., {Doyle}, J.~G., \& {Lemaire}, P. 2000, \solphys,
  194, 43

\bibitem[{{Banerjee} {et~al.}(1998){Banerjee}, {Teriaca}, {Doyle}, \&
  {Wilhelm}}]{Banerjee98}
{Banerjee}, D., {Teriaca}, L., {Doyle}, J.~G., \& {Wilhelm}, K. 1998, \aap,
  339, 208

\bibitem[{{Boland} {et~al.}(1975){Boland}, {Dyer}, {Firth}, {Gabriel}, {Jones},
  {Jordan}, {McWhirter}, {Monk}, \& {Turner}}]{Boland75}
{Boland}, B.~C., {Dyer}, E.~P., {Firth}, J.~G., {et~al.} 1975, \mnras, 171, 697

\bibitem[{{Boland} {et~al.}(1973){Boland}, {Engstrom}, {Jones}, \&
  {Wilson}}]{Boland73}
{Boland}, B.~C., {Engstrom}, S.~F.~T., {Jones}, B.~B., \& {Wilson}, R. 1973,
  \aap, 22, 161

\bibitem[{{Chae} {et~al.}(1998){Chae}, {Sch{\" u}hle}, \& {Lemaire}}]{Chae98}
{Chae}, J., {Sch{\" u}hle}, U., \& {Lemaire}, P. 1998, \apj, 505, 957

\bibitem[{{Contesse} {et~al.}(2004){Contesse}, {Koutchmy}, \&
  {Viladrich}}]{Contesse04}
{Contesse}, L., {Koutchmy}, S., \& {Viladrich}, C. 2004, Annales Geophysicae,
  22, 3055

\bibitem[{{Cranmer}(2000)}]{Cranmer00}
{Cranmer}, S.~R. 2000, \apj, 532, 1197

\bibitem[{{Cranmer} {et~al.}(1999{\natexlab{a}}){Cranmer}, {Field}, \&
  {Kohl}}]{Cranmer99b}
{Cranmer}, S.~R., {Field}, G.~B., \& {Kohl}, J.~L. 1999{\natexlab{a}}, \apj,
  518, 937

\bibitem[{{Cranmer} {et~al.}(1999{\natexlab{b}}){Cranmer}, {Kohl}, {Noci},
  {Antonucci}, {Tondello}, {Huber}, {Strachan}, {Panasyuk}, {Gardner},
  {Romoli}, {Fineschi}, {Dobrzycka}, {Raymond}, {Nicolosi}, {Siegmund},
  {Spadaro}, {Benna}, {Ciaravella}, {Giordano}, {Habbal}, {Karovska}, {Li},
  {Martin}, {Michels}, {Modigliani}, {Naletto}, {O'Neal}, {Pernechele},
  {Poletto}, {Smith}, \& {Suleiman}}]{Cranmer99}
{Cranmer}, S.~R., {Kohl}, J.~L., {Noci}, G., {et~al.} 1999{\natexlab{b}}, \apj,
  511, 481

\bibitem[{{David} {et~al.}(1998){David}, {Gabriel}, {Bely-Dubau}, {Fludra},
  {Lemaire}, \& {Wilhelm}}]{David98}
{David}, C., {Gabriel}, A.~H., {Bely-Dubau}, F., {et~al.} 1998, \aap, 336, L90

\bibitem[{{Dere} \& {Mason}(1993)}]{Dere93}
{Dere}, K.~P. \& {Mason}, H.~E. 1993, \solphys, 144, 217

\bibitem[{{Dolla}(2006)}]{Dolla_PhD}
{Dolla}, L. 2006, PhD thesis, Université Paris-Sud

\bibitem[{{Dolla} {et~al.}(2003){Dolla}, {Lemaire}, {Solomon}, \&
  {Vial}}]{Dolla03}
{Dolla}, L., {Lemaire}, P., {Solomon}, J., \& {Vial}, J.-C. 2003, in AIP Conf.
  Proc. 679: Solar Wind Ten, 351--354

\bibitem[{{Dolla} {et~al.}(2004){Dolla}, {Solomon}, \& {Lemaire}}]{Dolla04}
{Dolla}, L., {Solomon}, J., \& {Lemaire}, P. 2004, in ESA SP-547: SOHO 13
  Waves, Oscillations and Small-Scale Transients Events in the Solar
  Atmosphere: Joint View from SOHO and TRACE, 391

\bibitem[{{Doschek} \& {Feldman}(2000)}]{Doschek00}
{Doschek}, G.~A. \& {Feldman}, U. 2000, \apj, 529, 599

\bibitem[{{Doschek} {et~al.}(2001){Doschek}, {Feldman}, {Laming}, {Sch{\"
  u}hle}, \& {Wilhelm}}]{Doschek01}
{Doschek}, G.~A., {Feldman}, U., {Laming}, J.~M., {Sch{\" u}hle}, U., \&
  {Wilhelm}, K. 2001, \apj, 546, 559

\bibitem[{{Doschek} {et~al.}(1997){Doschek}, {Warren}, {Laming}, {Mariska},
  {Wilhelm}, {Lemaire}, {Schuehle}, \& {Moran}}]{Doschek97}
{Doschek}, G.~A., {Warren}, H.~P., {Laming}, J.~M., {et~al.} 1997, \apjl, 482,
  L109

\bibitem[{{Doyle} {et~al.}(1999){Doyle}, {Teriaca}, \& {Banerjee}}]{Doyle99}
{Doyle}, J.~G., {Teriaca}, L., \& {Banerjee}, D. 1999, \aap, 349, 956

\bibitem[{{Doyle} {et~al.}(2000){Doyle}, {Teriaca}, \& {Banerjee}}]{Doyle00}
{Doyle}, J.~G., {Teriaca}, L., \& {Banerjee}, D. 2000, \aap, 356, 335

\bibitem[{{Ekberg} \& {Feldman}(2003)}]{Ekberg03}
{Ekberg}, J.~O. \& {Feldman}, U. 2003, \apjs, 148, 567

\bibitem[{{Erdelyi} {et~al.}(1998){Erdelyi}, {Doyle}, {Perez}, \&
  {Wilhelm}}]{Erdelyi98}
{Erdelyi}, R., {Doyle}, J.~G., {Perez}, M.~E., \& {Wilhelm}, K. 1998, \aap,
  337, 287

\bibitem[{{Esser} {et~al.}(1999){Esser}, {Fineschi}, {Dobrzycka}, {Habbal},
  {Edgar}, {Raymond}, {Kohl}, \& {Guhathakurta}}]{Esser99}
{Esser}, R., {Fineschi}, S., {Dobrzycka}, D., {et~al.} 1999, \apjl, 510, L63

\bibitem[{{Feldman} {et~al.}(1997){Feldman}, {Behring}, {Curdt}, {Schuehle},
  {Wilhelm}, {Lemaire}, \& {Moran}}]{Feldman97}
{Feldman}, U., {Behring}, W.~E., {Curdt}, W., {et~al.} 1997, \apjs, 113, 195

\bibitem[{{Feldman} {et~al.}(2000){Feldman}, {Curdt}, {Landi}, \&
  {Wilhelm}}]{Feldman00}
{Feldman}, U., {Curdt}, W., {Landi}, E., \& {Wilhelm}, K. 2000, \apj, 544, 508

\bibitem[{{Feldman} {et~al.}(1999){Feldman}, {Doschek}, {Sch{\" u}hle}, \&
  {Wilhelm}}]{Feldman99}
{Feldman}, U., {Doschek}, G.~A., {Sch{\" u}hle}, U., \& {Wilhelm}, K. 1999,
  \apj, 518, 500

\bibitem[{{Frazin} {et~al.}(1999){Frazin}, {Modigliani}, {Ciaravella},
  {Dennis}, {Fineschi}, {Michels}, {Gardner}, {O'Neal}, {Raymond}, {Wu},
  {Noci}, \& {Kohl}}]{Frazin99}
{Frazin}, R.~A., {Modigliani}, A., {Ciaravella}, A., {et~al.} 1999, in American
  Institute of Physics Conference Series, 235

\bibitem[{{Gabriel} {et~al.}(2003){Gabriel}, {Bely-Dubau}, \&
  {Lemaire}}]{Gabriel03}
{Gabriel}, A.~H., {Bely-Dubau}, F., \& {Lemaire}, P. 2003, \apj, 589, 623

\bibitem[{{Habbal} {et~al.}(1993){Habbal}, {Esser}, \& {Arndt}}]{habbal93}
{Habbal}, S.~R., {Esser}, R., \& {Arndt}, M.~B. 1993, \apj, 413, 435

\bibitem[{{Hackenberg} {et~al.}(2000){Hackenberg}, {Marsch}, \&
  {Mann}}]{Hackenberg00}
{Hackenberg}, P., {Marsch}, E., \& {Mann}, G. 2000, \aap, 360, 1139

\bibitem[{{Harrison} {et~al.}(2002){Harrison}, {Hood}, \& {Pike}}]{Harrison02}
{Harrison}, R.~A., {Hood}, A.~W., \& {Pike}, C.~D. 2002, \aap, 392, 319

\bibitem[{{Hassler} {et~al.}(1990){Hassler}, {Rottman}, {Shoub}, \&
  {Holzer}}]{Hassler90}
{Hassler}, D.~M., {Rottman}, G.~J., {Shoub}, E.~C., \& {Holzer}, T.~E. 1990,
  \apjl, 348, L77

\bibitem[{{Hollweg} \& {Isenberg}(2002)}]{Hollweg02}
{Hollweg}, J.~V. \& {Isenberg}, P.~A. 2002, J. Geophys. Res., 12

\bibitem[{{Hu} {et~al.}(2000){Hu}, {Esser}, \& {Habbal}}]{Hu00}
{Hu}, Y.~Q., {Esser}, R., \& {Habbal}, S.~R. 2000, \jgr, 105, 5093

\bibitem[{{Isenberg}(2001)}]{Isenberg01b}
{Isenberg}, P.~A. 2001, \jgr, 29249

\bibitem[{{Isenberg} {et~al.}(2001){Isenberg}, {Lee}, \&
  {Hollweg}}]{Isenberg01}
{Isenberg}, P.~A., {Lee}, M.~A., \& {Hollweg}, J.~V. 2001, \jgr, 5649

\bibitem[{{Kohl} {et~al.}(1998){Kohl}, {Noci}, {Antonucci}, {Tondello},
  {Huber}, {Cranmer}, {Strachan}, {Panasyuk}, {Gardner}, {Romoli}, {Fineschi},
  {Dobrzycka}, {Raymond}, {Nicolosi}, {Siegmund}, {Spadaro}, {Benna},
  {Ciaravella}, {Giordano}, {Habbal}, {Karovska}, {Li}, {Martin}, {Michels},
  {Modigliani}, {Naletto}, {O'Neal}, {Pernechele}, {Poletto}, {Smith}, \&
  {Suleiman}}]{Kohl98}
{Kohl}, J.~L., {Noci}, G., {Antonucci}, E., {et~al.} 1998, \apjl, 501, L127

\bibitem[{{Kopp} \& {Holzer}(1976)}]{Kopp76}
{Kopp}, R.~A. \& {Holzer}, T.~E. 1976, \solphys, 49, 43

\bibitem[{{Lee} {et~al.}(2000){Lee}, {Yun}, \& {Chae}}]{Lee00}
{Lee}, H., {Yun}, H.~S., \& {Chae}, J. 2000, Journal of Korean Astronomical
  Society, 33, 57

\bibitem[{{Lemaire} {et~al.}(1997){Lemaire}, {Wilhelm}, {Curdt}, {Schule},
  {Marsch}, {Poland}, {Jordan}, {Thomas}, {Hassler}, {Vial}, {Kuhne}, {Huber},
  {Siegmund}, {Gabriel}, {Timothy}, \& {Grewing}}]{Lemaire97}
{Lemaire}, P., {Wilhelm}, K., {Curdt}, W., {et~al.} 1997, \solphys, 170, 105

\bibitem[{{Li}(2002)}]{Li02}
{Li}, X. 2002, \apjl, 571, L67

\bibitem[{{Li} {et~al.}(1998){Li}, {Habbal}, {Kohl}, \& {Noci}}]{Li98}
{Li}, X., {Habbal}, S.~R., {Kohl}, J., \& {Noci}, G. 1998, \apjl, 501, L133

\bibitem[{{Liewer} {et~al.}(1999){Liewer}, {Velli}, \& {Goldstein}}]{Liewer99}
{Liewer}, P., {Velli}, M., \& {Goldstein}, B. 1999, Space Science Reviews, 87,
  257

\bibitem[{{Mariska} {et~al.}(1978){Mariska}, {Feldman}, \&
  {Doschek}}]{Mariska78}
{Mariska}, J.~T., {Feldman}, U., \& {Doschek}, G.~A. 1978, \apj, 226, 698

\bibitem[{{Markovskii}(2001)}]{Markovskii01}
{Markovskii}, S.~A. 2001, \apj, 557, 337

\bibitem[{{Markovskii} \& {Hollweg}(2002)}]{Markovskii02}
{Markovskii}, S.~A. \& {Hollweg}, J.~V. 2002, \grl, 29, 24

\bibitem[{{Marsch} \& {Tu}(2001)}]{Marsch01}
{Marsch}, E. \& {Tu}, C.-Y. 2001, \jgr, 106, 227

\bibitem[{{Moran}(2001)}]{Moran01}
{Moran}, T.~G. 2001, \aap, 374, L9

\bibitem[{{Moran}(2003)}]{Moran03}
{Moran}, T.~G. 2003, \apj, 598, 657

\bibitem[{{Morgan} \& {Habbal}(2004)}]{Morgan04b}
{Morgan}, H. \& {Habbal}, S.~R. 2004, in Bulletin of the American Astronomical
  Society, Vol.~36, Bulletin of the American Astronomical Society, 695

\bibitem[{{Ofman}(2004)}]{Ofman04}
{Ofman}, L. 2004, in ESA SP-547: SOHO 13 Waves, Oscillations and Small-Scale
  Transients Events in the Solar Atmosphere: Joint View from SOHO and TRACE,
  345

\bibitem[{{O'Shea} {et~al.}(2004){O'Shea}, {Banerjee}, \& {Doyle}}]{O'Shea04}
{O'Shea}, E., {Banerjee}, D., \& {Doyle}, G. 2004, in ESA SP-575: SOHO 15
  Coronal Heating, 142

\bibitem[{{O'Shea} {et~al.}(2003){O'Shea}, {Banerjee}, \& {Poedts}}]{O'shea03}
{O'Shea}, E., {Banerjee}, D., \& {Poedts}, S. 2003, \aap, 400, 1065

\bibitem[{{Parker}(1991)}]{Parker91}
{Parker}, E.~N. 1991, \apj, 372, 719

\bibitem[{{Patsourakos} {et~al.}(2002){Patsourakos}, {Habbal}, \&
  {Hu}}]{Patsourakos02}
{Patsourakos}, S., {Habbal}, S.~R., \& {Hu}, Y.~Q. 2002, \apjl, 581, L125

\bibitem[{{Pek{\" u}nl{\" u}} {et~al.}(2002){Pek{\" u}nl{\" u}}, {Bozkurt},
  {Afsar}, {Soydugan}, \& {Soydugan}}]{Pekunlu02}
{Pek{\" u}nl{\" u}}, E.~R., {Bozkurt}, Z., {Afsar}, M., {Soydugan}, E., \&
  {Soydugan}, F. 2002, \mnras, 336, 1195

\bibitem[{{Peter} \& {Vocks}(2003)}]{Peter03}
{Peter}, H. \& {Vocks}, C. 2003, \aap, 411, L481

\bibitem[{{Schwenn} \& {Marsch}(1991)}]{Schwenn91}
{Schwenn}, R. \& {Marsch}, E. 1991, Physics of the {I}nner {H}eliosphere {II},
  {P}articules, Waves and Turbulence (Springer-Verlag)

\bibitem[{{Seely} {et~al.}(1997){Seely}, {Feldman}, {Schuehle}, {Wilhelm},
  {Curdt}, \& {Lemaire}}]{Seely97}
{Seely}, J.~F., {Feldman}, U., {Schuehle}, U., {et~al.} 1997, \apjl, 484, L87

\bibitem[{{Singh} {et~al.}(2003{\natexlab{a}}){Singh}, {Ichimoto}, {Sakurai},
  \& {Muneer}}]{Singh03}
{Singh}, J., {Ichimoto}, K., {Sakurai}, T., \& {Muneer}, S. 2003{\natexlab{a}},
  \apj, 585, 516

\bibitem[{{Singh} {et~al.}(2003{\natexlab{b}}){Singh}, {Sakurai}, {Ichimoto},
  \& {Muneer}}]{Singh03b}
{Singh}, J., {Sakurai}, T., {Ichimoto}, K., \& {Muneer}, S. 2003{\natexlab{b}},
  \solphys, 212, 343

\bibitem[{{Tu} \& {Marsch}(1997)}]{Tu97}
{Tu}, C.~Y. \& {Marsch}, E. 1997, \solphys, 171, 363

\bibitem[{{Tu} {et~al.}(1998){Tu}, {Marsch}, {Wilhelm}, \& {Curdt}}]{Tu98}
{Tu}, C.-Y., {Marsch}, E., {Wilhelm}, K., \& {Curdt}, W. 1998, \apj, 503, 475

\bibitem[{{Vocks} \& {Marsch}(2002)}]{Vocks02b}
{Vocks}, C. \& {Marsch}, E. 2002, \apj, 568, 1030

\bibitem[{{Wesson}(2004)}]{Wesson04}
{Wesson}, J.~A. 2004, {Tokamaks, third edition} (Oxford Science Pub.)

\bibitem[{{Wilhelm} {et~al.}(1995){Wilhelm}, {Curdt}, {Marsch}, {Schuhle},
  {Lemaire}, {Gabriel}, {Vial}, {Grewing}, {Huber}, {Jordan}, {Poland},
  {Thomas}, {Kuhne}, {Timothy}, {Hassler}, \& {Siegmund}}]{Wilhelm95}
{Wilhelm}, K., {Curdt}, W., {Marsch}, E., {et~al.} 1995, \solphys, 162, 189

\bibitem[{{Wilhelm} {et~al.}(2004){Wilhelm}, {Dwivedi}, \&
  {Teriaca}}]{Wilhelm04}
{Wilhelm}, K., {Dwivedi}, B.~N., \& {Teriaca}, L. 2004, \aap, 415, 1133

\bibitem[{{Wilhelm} {et~al.}(2005){Wilhelm}, {Fludra}, {Teriaca}, {Harrison},
  {Dwivedi}, \& {Pike}}]{Wilhelm05}
{Wilhelm}, K., {Fludra}, A., {Teriaca}, L., {et~al.} 2005, \aap, 435, 733

\bibitem[{{Wilhelm} {et~al.}(1997){Wilhelm}, {Lemaire}, {Curdt}, {Schuhle},
  {Marsch}, {Poland}, {Jordan}, {Thomas}, {Hassler}, {Huber}, {Vial}, {Kuhne},
  {Siegmund}, {Gabriel}, {Timothy}, {Grewing}, {Feldman}, {Hollandt}, \&
  {Brekke}}]{Wilhelm97}
{Wilhelm}, K., {Lemaire}, P., {Curdt}, W., {et~al.} 1997, \solphys, 170, 75

\bibitem[{{Wilhelm} {et~al.}(1998){Wilhelm}, {Marsch}, {Dwivedi}, {Hassler},
  {Lemaire}, {Gabriel}, \& {Huber}}]{Wilhelm98}
{Wilhelm}, K., {Marsch}, E., {Dwivedi}, B.~N., {et~al.} 1998, \apj, 500, 1023

\end{thebibliography}

\end{document}